\documentclass[sigconf]{acmart}

\copyrightyear{2025}
\acmYear{2025}
\setcopyright{acmlicensed}\acmConference[CHI '25]{CHI Conference on Human Factors in Computing Systems}{April 26-May 1, 2025}{Yokohama, Japan}
\acmBooktitle{CHI Conference on Human Factors in Computing Systems (CHI '25), April 26-May 1, 2025, Yokohama, Japan}
\acmDOI{10.1145/3706598.3713862}
\acmISBN{979-8-4007-1394-1/25/04}

\usepackage{acmart-taps}

\usepackage{booktabs} 

\usepackage{enumitem}
\setlist[itemize]{leftmargin=*}
\setlist[enumerate]{leftmargin=*,label=\arabic*.}

\usepackage[shortcuts]{extdash}

\usepackage{soul}

\usepackage{xcolor}

\usepackage{eurosym}

\usepackage{xspace}

\usepackage{xparse}

\newcommand*{\icon}[1]{%
    \raisebox{-.15\baselineskip}{%
        \includegraphics[
        height=0.8\baselineskip,
        width=0.8\baselineskip,
        keepaspectratio,
        ]{#1}%
    }%
}

\newcount\S \newcount\H  \newcount\Hrest  \newcount\M  \newcount\Mrest
\newcommand{\prettytimestamp}[1]{%
   \setS#1.\end
   \H=\S \divide\H by3600
   \Hrest=\H \multiply\Hrest by-3600 \advance\Hrest by\S
   \M=\Hrest \divide\M by60
   \Mrest=\M \multiply\Mrest by-60 \advance\Mrest by\Hrest
   \ifnum \H=0
     \ifnum\M=0 \miliseconds#1s...\end             
     \else \the\M min \the\Mrest s\fi            
   \else \the\H h\the\M min\the\Mrest s\fi 
}
\def\setS#1.#2\end{\S=#1\relax}
\def\miliseconds#1.#2#3#4\end{#1\ifx.#2\else.#2\ifx.#3\else#3\fi\fi}

\usepackage[nameinlink]{cleveref}
\usepackage{stfloats}

\usepackage{booktabs}

\newcommand{\name}{Textoshop\xspace}

\begin{document}

\tolerance=400 

%
\title{\name: Interactions Inspired by Drawing Software to Facilitate Text Editing}

\author{Damien Masson}
\orcid{0000-0002-9482-8639}
\affiliation{%
  \institution{University of Toronto}
  \department{Department of Computer Science}
  \city{Toronto}
  \country{Canada}
}
\additionalaffiliation{\institution{Université de Montréal}
  \city{Montreal}
  \country{Canada}
}
\email{damien.masson@umontreal.ca}

\author{Young-Ho Kim}
\orcid{0000-0002-2681-2774}
\affiliation{%
  \institution{NAVER AI Lab}
  \city{Seongnam}
  \country{Republic of Korea}
}
\email{yghokim@younghokim.net}

\author{Fanny Chevalier}
\orcid{0000-0002-5585-7971}
\affiliation{%
  \institution{University of Toronto}
  \department{Department of Computer Science and
Statistical Sciences}
  \city{Toronto}
  \country{Canada}
}
\email{fanny@cs.toronto.edu}

\begin{abstract}
We explore how interactions inspired by drawing software can help edit text. Making an analogy between visual and text editing, we consider words as pixels, sentences as regions, and tones as colours. For instance, direct manipulations move, shorten, expand, and reorder text; tools change number, tense, and grammar; colours map to tones explored along three dimensions in a tone picker; and layers help organize and version text. This analogy also leads to new workflows, such as boolean operations on text fragments to construct more elaborated text. A study shows participants were more successful at editing text and preferred using the proposed interface over existing solutions. Broadly, our work highlights the potential of interaction analogies to rethink existing workflows, while capitalizing on familiar features.
\end{abstract}

%
%



 \begin{teaserfigure}
   \includegraphics[width=\textwidth]{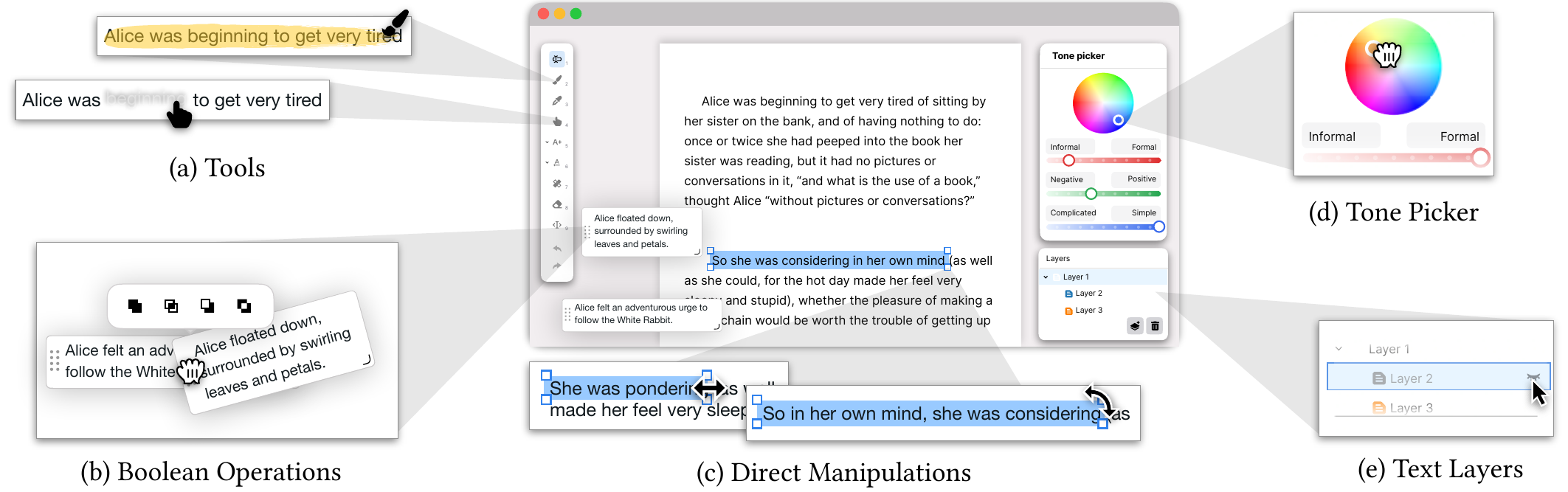}
   \caption{\name uses interactions inspired by drawing software to help edit text: (a) text can be revised using tools such as the smudge and tone brush; (b) text passages can be merged using boolean operations; (c) text can be shortened and reordered through direct manipulations; (d) tones can be explored in a tone picker; and (e) text can be organized in layers}
   \Description{Screenshot of the Textoshop interface with call-outs to show each feature independently. Tools are demonstrated by showing the effect of using the tone brush and smudger. Boolean operations are demonstrated by dragging two text passages. Direct manipulation is demonstrated by showing the effect of resizing and rotating a selection. Text layers are demonstrated by hiding a layer. The tone picker is demonstrated by moving the thumb in the colour wheel.}
   \label{fig:teaser}
\end{teaserfigure}

\sloppy
\maketitle



\section{Introduction}
The processes involved in writing share many similarities with those involved in image editing: words form sentences to tell a story, much like pixels come together to create regions that form a complete picture.
Just as writers plan and refine a story by outlining the narrative and rearranging the story flow~\cite{flowerProblemSolvingStrategiesWriting1977}, 
visual artists plan and refine a graphical composition by laying out visual elements on a canvas, rearranging them, sometimes organizing them in layers ~\cite{tsandilasBricoSketchMixingPaper2015}.
Just as writers combine arguments into sentences and paragraphs~\cite{oshimaWritingAcademicEnglish2005}, visual artists combine shapes into complex drawings~\cite{woodAdobeIllustratorClassroom2022}.
And just as writers revise the language and iterate on the wording and structure~\cite{faigleyAnalyzingRevision1981}, visual artists revise the aesthetics and iterate on the colours and textures~\cite{tsandilasBricoSketchMixingPaper2015}.

Despite similarities in \emph{process}, the interaction paradigm to support digital writing differs substantially from the powerful ways to manipulate visual content.
For example, in image editing and vector graphics software like Adobe Photoshop and Illustrator, artists can plan by directly manipulating elements of the image, moving them around and outside the canvas, and organizing them in independent layers. 
They can compose complex geometric structures by uniting, intersecting, or subtracting simple geometries. 
And they can rapidly explore nuances by navigating the space of colours in a colour picker. 
Even a task as simple as resizing an element, which drawing software support by dragging the selection box handles, will require careful rewriting with a text editor---think about the amount of effort involved in wordsmithing a paragraph to make it exactly fit within the space left on a page.

While large language models (LLMs) can help with writing tasks, how we interact with them rarely achieves the simplicity and expressiveness of drawing software. For example, LLMs must be prompted using relatively slow text input, preventing quick and precise iterations. These prompts are often verbose, and users may struggle to convey their intent using only words~\cite{massonDirectGPTDirectManipulation2024, chungPromptPaintSteeringTexttoImage2023}. In contrast, drawing software rely on direct manipulation of graphical elements. They also offer tools for smart edits. For example, the colour picker allows for exploring the space of solutions and expressing intents that are difficult to formulate in words. We see an opportunity to learn from this interaction paradigm to rethink the digital writing workflow.

To better support writers\footnote{By ``writers'', we do not necessarily mean professional writers but any individual who engages in writing activities, no matter their expertise}, we propose to adapt solutions used in drawing software to help with recurring writing tasks. Specifically, we aim to (1) capitalize on solutions that have been proven to work well and (2) offer solutions most users are already familiar with.
Loosely, we view \textit{words} as \textit{pixels}, \textit{sentences} as \textit{regions}, and \textit{tones} as \textit{colours}. 
This analogy resulted in the design of \name (\Cref{fig:teaser}), a text editing software that supports meaning-preserving changes through direct manipulation to move, resize and restructure text selections; a layer management system to organize, display, and edit text in independent views; a sentence composition feature to construct sentences or paragraphs by combining, intersecting, or subtracting ideas or phrases; a tone picker to explore different tones when editing text; and tools including an ``eraser'' to intelligently remove part of the text, a ``tone eyedropper'' to copy the tone of a passage, and a ``smudge'' tool to paraphrase and smoothen transitions between sentences.

In a user study with 12 lay individuals who frequently engage in writing activities, \name was compared to existing solutions when accomplishing text editing tasks such as organizing and integrating sentence fragments, changing the tone, using a template, and summarizing and expanding a passage. In a second part, participants used \name in a free-form exploratory task. Results show that participants were more successful and preferred using \name over alternatives. Participants appreciated the design of \name and understood the analogy, allowing them to transfer some of their knowledge of drawing software to use the tool. Overall, our work promotes leveraging interaction analogies to reimagine existing workflows.

\section{Background and Related Work}
Traditional text editors such as Microsoft Word offer limited support for editing the text. We first review the issues this creates before discussing solutions proposed in the literature. We then detail the rationale for taking inspiration from drawing software\footnote{We use the term ``drawing software'' to refer to both raster (e.g., Adobe Photoshop) and vector (e.g., Inkscape and Adobe Illustrator) graphics editors. }.

\subsection{Issues with Traditional Ways of Editing Text}\label{sec:issues}
First, \textit{organization capabilities are limited when relying solely on the metaphor of sheets of paper}. Writers go through different stages in a non-linear order~\cite{flowerCognitiveProcessTheory1981}. For example, they often alternate between generating ideas, constructing sentences, and revising these sentences~\cite{chakrabartyCreativitySupportAge2024}. In fact, it is often recommended for writers to organize ideas into spatial representations~\cite{flowerProblemSolvingStrategiesWriting1977, shibataCognitiveSupportOrganization2008}. Yet, text editors force organization in lines on pages, often leading to messy drafts mixing ideas, outlines, revisions, and discarded alternatives. This often results in a poor organization limiting writers' exploration~\cite{rezaABScribeRapidExploration2024}.

Second, \textit{exploring the space of possible text variations is difficult}. A sentence can be rewritten in many ways by varying tone, word order, or length. In fact, rewriting is often considered ``the essence of writing''~\cite{zinsserWritingWellClassic2016}. Yet, most text editors offer little support to help revise language other than a thesaurus to find synonyms. Some experimental tools leverage LLMs to revise text, but it remains challenging for writers to express their intent with fine granularity~\cite{kimUnderstandingUsersDissatisfaction2024a}.

Third, \textit{reorganizing ideas already expressed is tedious} because ideas are difficult to isolate and text fragments are difficult to combine. Once ideas are expressed into sentences and paragraphs, moving these ideas requires rewriting most of the sentence or paragraph to guarantee consistency~\cite{kingReverseOutliningMethod2012}. Besides making the exploration of alternatives difficult, it may also cause writer's block as it reinforces this view of an immutable text structure~\cite{roseRigidRulesInflexible1980}.

Fourth, \textit{managing and rendering different versions of a text is tedious}. Most text editors consider text as a linear sequence of characters with no structure other than chapters, sections, and paragraphs. Besides, edits are typically destructive, with only support for sequentially reversing the most recent changes incrementally with ``undo''', or lookup in document histories that are tedious to browse~\cite{chevalierUsingTextAnimated2010}. This means that finding specific content and storing alternatives might be challenging. Writers may need to split their writing into different files or keep different copies of their text for each version, which is tedious to manage~\cite{hendersonDocumentDuplicationHow2011}.

\subsection{Novel Ways of Editing Text}
Previous work has proposed solutions to some of the above issues, often using generative AI. Below, we describe these solutions and categorize them based on the problem they address.

\subsubsection{Writing Using Non-linear Text Editors} 
As early as the development of the World Wide Web, \citet{bolterHypertextCreativeWriting1987} proposed a writing tool supporting interactive fiction writing using a structural editor with a diagrammatic view. This spurred research into the development of hypertext authoring tools~\cite{kitromiliWhatHypertextAuthoring2019} with which authors write part of the story in an order that might differ from how readers will read it. While these tools were primarily designed to create interactive stories, the idea of being able to write in any order inspired tools to author more traditional and static text. For example, the commercial software Scrivener~\cite{Scrivener} allows writers to compose text in any order by relying on different text areas that can be reorganized to rearrange the final story.

\name also supports non-linear writing: text can be either positioned following lines on a page or freely arranged in 2D in and off the page.
The margins can store mood boards, alternatives, and outlines akin to how digital artists leverage the space outside the canvas to preserve elements for reference~\cite{holinatySupportingReferenceImagery2021}. 
Additionally, \name supports layers of text that can be edited independently and combined at will to create different composite stories.

\subsubsection{Revising and Exploring Text Variations} 
While LLMs can generate text variations, steering the generation remains challenging. In its simplest form, previous work has proposed to select a passage and then prompt the model~\cite{massonDirectGPTDirectManipulation2024, hoqueHaLLMarkEffectSupporting2024} or click a button~\cite{rezaABScribeRapidExploration2024} to generate text alternatives. However, this interaction gives little freedom to control the result. Instead, other work proposed to generate not one but thousands of text variations in the hope that one of them might be satisfactory. For example, \citet{geroSupportingSensemakingLarge2024} explored different ways to present and query these different versions. Similarly, Luminate~\cite{suhLuminateStructuredGeneration2024} arranges the different variations along specific dimensions, making it easier for users to find a variation of interest.

Our work differs in that the space of possible variations is explored through direct manipulation. If a sentence is close to satisfactory, users can explore variations of different lengths or word orders through the resizing and rotating features. They can also move a slider in a two-dimensional space to explore tone nuances, much like visual artists explore colour nuances in a colour picker.

\subsubsection{Reordering and Combining Text Fragments} New writing tools increasingly \textit{``shift the writer's role towards editorial decisions''} and manipulating text fragments~\cite{buschekCollageNewWriting2024}. 
For example, Langsmith~\cite{itoLangsmithInteractiveAcademic2020} can suggest full sentences from incomplete and rough sentence fragments. Similarly, Rambler~\cite{linRamblerSupportingWriting2024} allows refining and combining dictated train-of-thoughts into full drafts. Other work proposed to manipulate summary representations such as paragraph summaries~\cite{dangTextGenerationSupporting2022a} and visual outlines~\cite{zhangVISARHumanAIArgumentative2023} to revise the flow of ideas.

We go beyond simple merging of text passages by exploring novel ways to combine text inspired by boolean operations on shapes, such as subtracting to remove the information in one passage from another, intersecting to preserve only shared information between two passages, and excluding to remove redundant information.

\subsubsection{Text Versioning} Most text editors have some form of version control, albeit it is often limited. For example, Microsoft Word can track changes made to a document, and Google Docs and Scrivener~\cite{Scrivener} support storing, restoring, and comparing snapshots. More advanced versioning systems include ABScribe~\cite{rezaABScribeRapidExploration2024}, which allows storing multiple text variations at specific text locations. 

Our work explores the use of text layers that can also store modifications at specific text locations. Further, these layers can also store spatially disjoint but semantically related edits. This allows supporting new workflows where writers store alternatives and hide or show layers to try out different versions. And this also supports new use cases such as templating.

\subsection{Designing Interactions Based on Analogies}
At its core, our approach repurposes interactions from one domain (in our case, image editing) to another domain (i.e. text editing). This capitalizes on features familiar to users and might inspire new workflows and new ways to think about the application domain. 

Designing software interactions based on such analogies (also referred to as interface metaphors~\cite{nealeChapter20Role1997}) can lead to easily learned and expressive interactions~\cite[ch 2.4]{rogersInteractionDesignHumanComputer2023}. For example, the direct manipulation interaction paradigm is all about manipulating elements in software as if they were physical objects that can be moved, resized, or distorted~\cite{shneidermanDirectManipulationStep1983, hutchinsDirectManipulationInterfaces1985}.
Instrumental interaction extends direct manipulation by noting that humans leverage tools when the finger or the hand is not enough~\cite{beaudouin-lafonInstrumentalInteractionInteraction2000, beaudouin-lafonReificationPolymorphismReuse2000}. In the software domain, these tools are interaction instruments that operate on domain objects.

Drawing software combine a canvas analogy with ideas from direct and instrumental interaction.
For example, Adobe Photoshop allows elements to be moved, removed, resized, and rotated through rapid and reversible physical actions that mimic how one interacts with physical objects in the real world, just like advocated by direct manipulation. 
And interaction instruments simulate paint brushes, pencils, or erasers. Other interaction instruments include the colour picker, which is analogous to a paint palette, and layers, which are analogous to transparent sheets~\cite{belgeBackFutureGraphical1993, andrewsAdobePhotoshopElements2006} and \textit{reify}~\cite{beaudouin-lafonReificationPolymorphismReuse2000} the concept of independent manipulation.

The success of drawing software has been inspirational for other software. For example, it is common to find the concept of brushes, colour pickers, and layers adopted by software operating in different domains, such as presentation and video editing. And some of the digital drawing concepts have been adapted successfully, such as using a paintbrush to control the fortune of characters in a story~\cite{chungTaleBrushSketchingStories2022}, mixing prompts just like colours can be mixed to convey visual styles difficult to express with words~\cite{chungPromptPaintSteeringTexttoImage2023}, or repurposing raster image tools to manipulate colour histograms~\cite{chevalierHistomagesFullySynchronized2012}.
Beyond offering more expressive interactions, these analogies help people reason about unfamiliar software by leveraging their accumulated interaction knowledge about digital tools~\cite{renomInteractionKnowledgeUnderstanding2023}.

In contrast, text editors evolved from command-line word processors~\cite{mahlowWritingToolsLooking2023, haighRememberingOfficeFuture2006} and use the analogy of the typewriter and sheet of paper~\cite{kirschenbaumTrackChangesLiterary2016}. In this work, we explore how interaction inspired by the graphical domain could open up new possibilities for writing.

\section{Textoshop}\label{sec:textoshop}
We reimagine the design of text editors in \name, a text editor that incorporates features and workflows borrowed from digital graphics editing.
\name employs a digital canvas instead of the sheet of paper metaphor. This means that text fragments are not seen as ink on a page but more like pieces of a collage that can be rearranged, resized, and modified at any time. Much like digital drawing software, \name also offers smart tools to manipulate and modify these text fragments.

We designed \name by reviewing features found in most raster or vector image editing software and that computer users are likely familiar with, such as colour pickers, drawing tools, and layers. We then mapped these features to novel text editing capabilities, favouring fast interactions for frequent text editing tasks as identified in the taxonomy from \citet{faigleyAnalyzingRevision1981}.
Below, we categorize these features based on this taxonomy, which differentiates between formal changes (e.g., spelling, tense, number), meaning-preserving changes (e.g., substituting a word with a synonym, reordering words in a sentence), and micro/macro changes.


\subsection{Dragging, Resizing, Rotating, and Splitting}
Direct interactions are typically reserved for frequent operations. In \name, we map these interactions to meaning-preserving changes whose impact is limited to the selection, such as additions, deletions, permutations, distributions, and consolidations~\cite{faigleyAnalyzingRevision1981}.

\begin{figure*}[htb]
\centering
    \includegraphics[width=\textwidth]{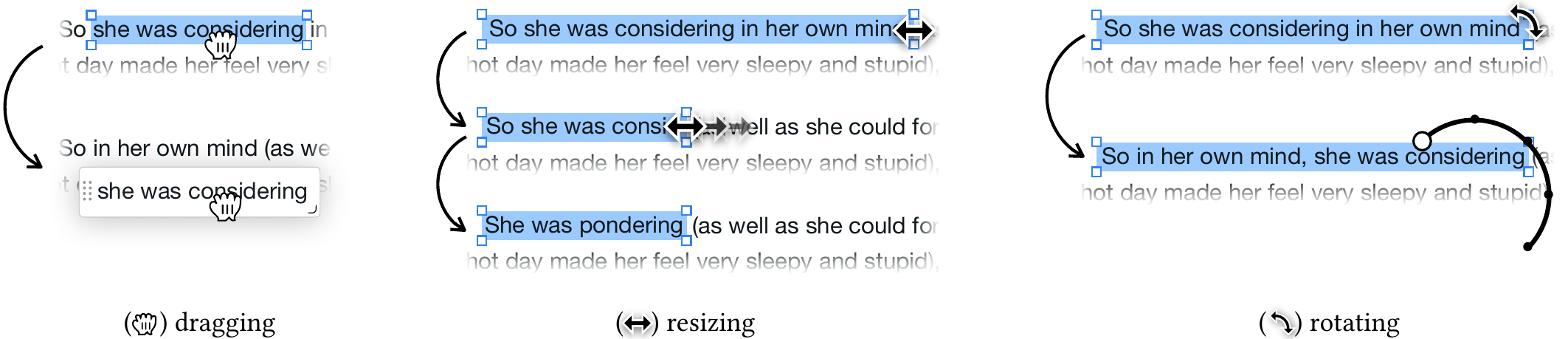}
\caption{Direct manipulations on text selections edit the text: (\raisebox{-2pt}{\protect\includegraphics[width=9pt]{figures/tools/dragging}}) dragging moves the selection anywhere on the canvas; (\begin{imageonly}\icon{figures/tools/resizing}\end{imageonly}) resizing shortens or expands the text; (\begin{imageonly}\icon{figures/tools/rotating}\end{imageonly}) rotating changes the order of the words.}
    \label{fig:directManipulations}
    \Description{Screenshots of the different direct manipulation interactions. Dragging moves the text selection. Resizing is demonstrated in the sentence "So she was considering in her own mind," which becomes "She was pondering". Rotating is demonstrated in the same sentence and results in "So in her own mind, she was considering".}
\end{figure*}

\medskip\noindent\begin{imageonly}\icon{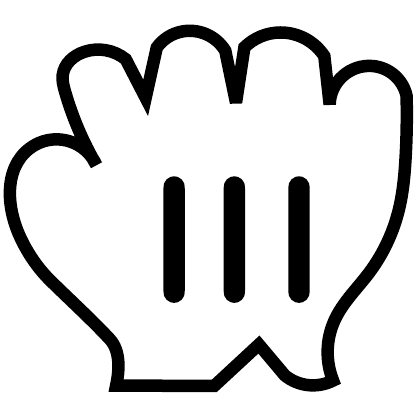}\end{imageonly} \textit{Dragging --} Freely moves a text selection in and off the the page through a drag-and-drop interaction (\Cref{fig:directManipulations}.\aptLtoX[graphic=no,type=html]{\raisebox{-2pt}{\includegraphics[width=9pt]{figures/tools/dragging.pdf}}}{\icon{figures/tools/dragging.pdf}}). Once dragged, the selected text is freed from the positioning constraints of the page and can be placed anywhere on or around the page on the canvas, supporting organizations similar to drawing software.

\medskip\noindent\aptLtoX[graphic=no,type=html]{\raisebox{-2pt}{\includegraphics[width=9pt]{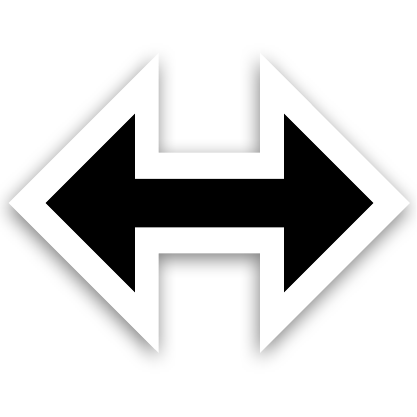}}}{\icon{figures/tools/resizing.pdf}} \textit{ Resizing --} Lengthens or shortens the selected text to fit within the specified space. The interaction is triggered by starting a dragging motion from the right edge of the text (\Cref{fig:directManipulations}.\aptLtoX[graphic=no,type=html]{\raisebox{-2pt}{\includegraphics[width=9pt]{figures/tools/resizing.pdf}}}{\icon{figures/tools/resizing.pdf}}). The modification is meaning-preserving. As such, shortening often removes words, while lengthening adds words non-essential to the core meaning of the text. Large expansions of a text selection might result in added descriptions and new sentences.

\medskip\noindent\aptLtoX[graphic=no,type=html]{\raisebox{-2pt}{\includegraphics[width=9pt]{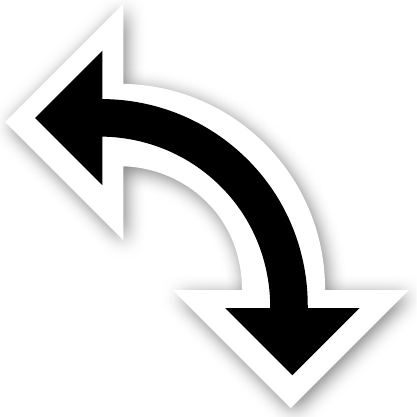}}}{\icon{figures/tools/rotating.pdf}} \textit{ Rotating --} Reorders the words in the selected text. The interaction is activated through a dragging motion starting on one of the four corner handles. The further the rotation thumb is moved from the center, the further the words are moved away from their original position (\Cref{fig:directManipulations}.\aptLtoX[graphic=no,type=html]{\raisebox{-2pt}{\includegraphics[width=9pt]{figures/tools/rotating.pdf}}}{\icon{figures/tools/rotating.pdf}}). For example, rotating ``The ball was kicked by Lisa'' reorders the words by using the active instead of the passive voice: ``Lisa kicked the ball''. 

\medskip\noindent\begin{imageonly}\icon{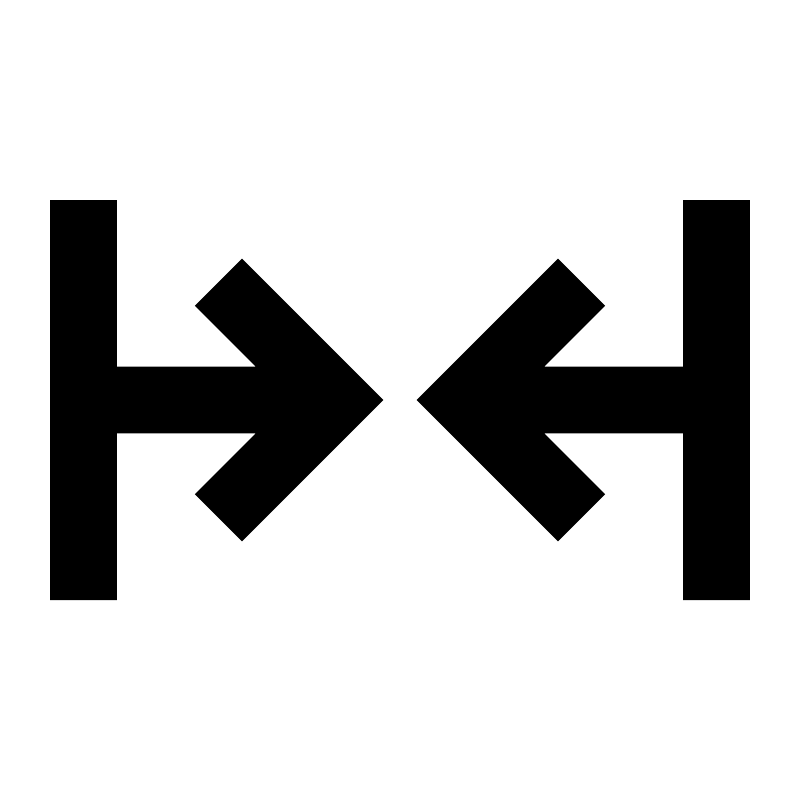}\end{imageonly} \textit{Combining and }\begin{imageonly}\icon{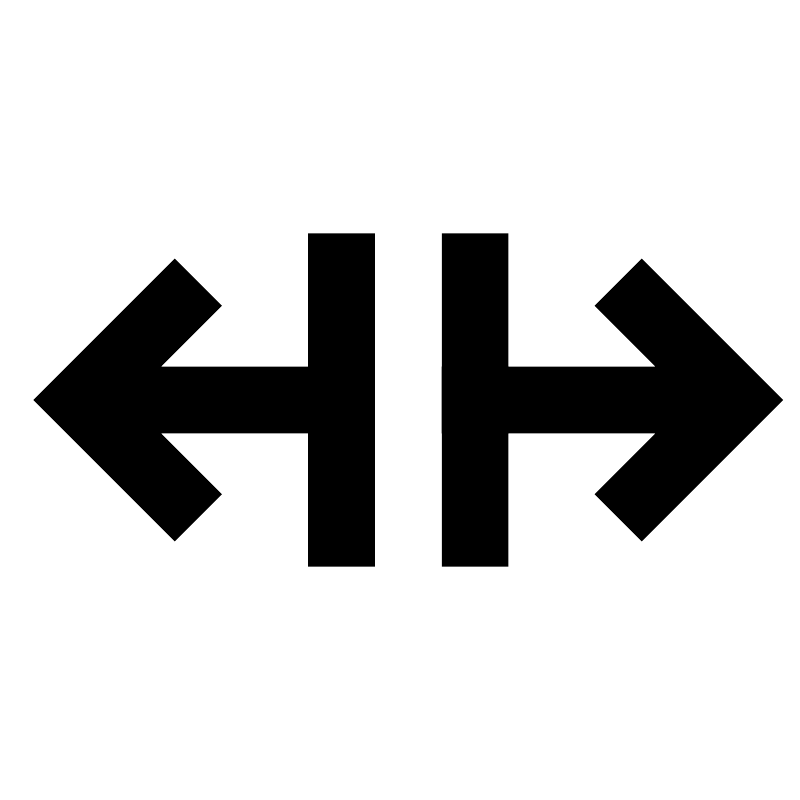}\end{imageonly} \textit{Splitting Sentences --} Splits a sentence into multiple, or consolidates multiple sentences into one. The action is in a contextual menu opened by right-clicking a text selection. 

\subsection{Tools}
Tools extend direct manipulation capabilities by activating modes that change the meaning of direct interactions~\cite{beaudouin-lafonInstrumentalInteractionInteraction2000, beaudouin-lafonReificationPolymorphismReuse2000}. \name maps tools to modifications that are meaning-preserving but whose impact can go beyond the selection.

\medskip\noindent\begin{imageonly}\icon{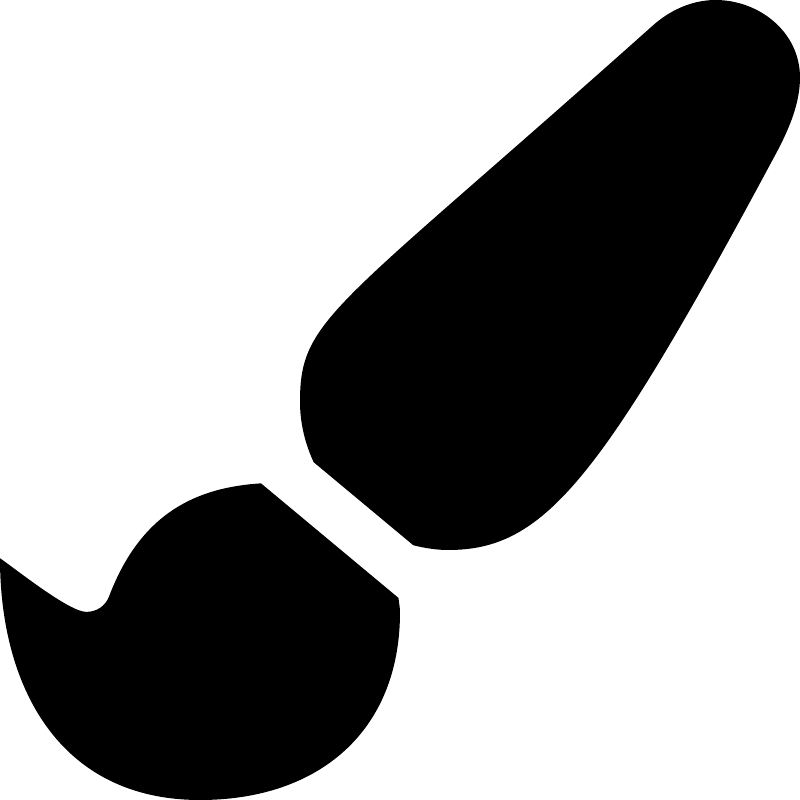}\end{imageonly} \textit{Tone Brush --} \label{sec:toneBrush}Changes the tone of the selected text using the currently selected tone (in the tone picker, \cref{sec:tonePicker}). For example, brushing the text ``Alice was beginning to get very tired'' with an informal tone would result in ``Alice was \textit{totally wiped out}''.

\medskip\noindent\begin{imageonly}\icon{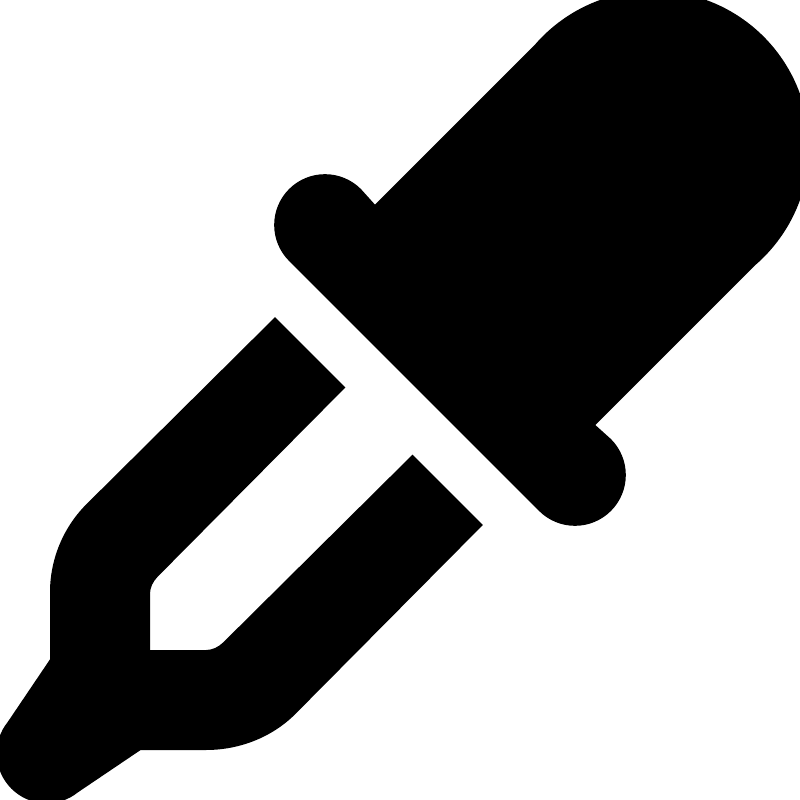}\end{imageonly} \textit{Tone Eyedropper --} Copies the tone of the selected text and update the tone picker (\cref{sec:tonePicker}). For example, eyedropping the text ``Alice was utterly exhausted'' updates the tone picker to a tone mostly negative, slightly formal, and complicated. The copied tone can then be applied to another part of the text using the tone brush.

\medskip\noindent\begin{imageonly}\icon{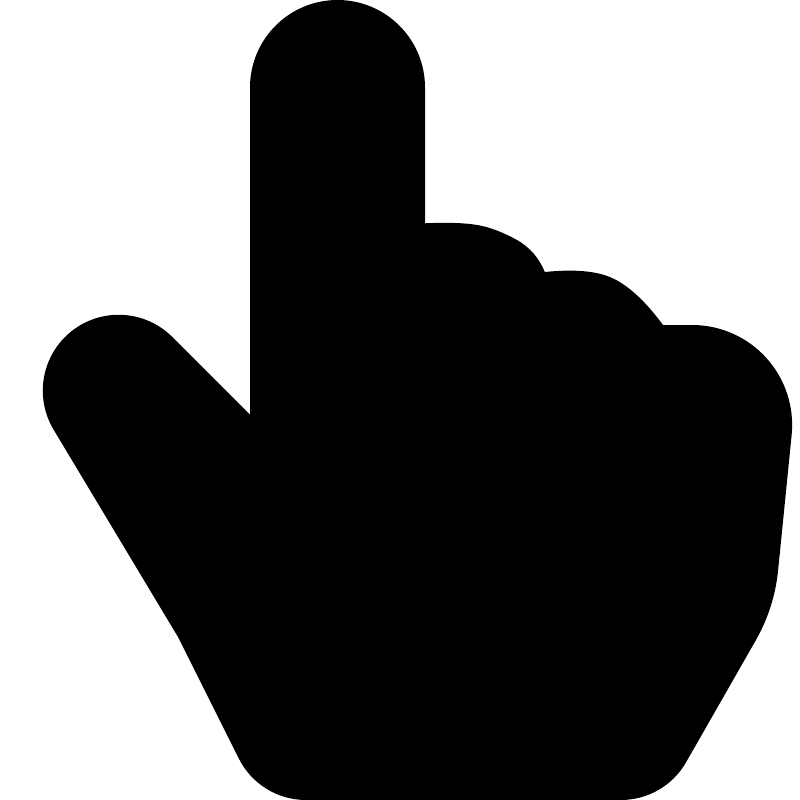}\end{imageonly} \textit{Smudge --} Regenerates the selected text. For example, smudging the word ``beginning'' modifies it to ``starting''. Smudging can also help connect or change the transition between two sentences by smudging the end of a sentence and the start of the next one.

\medskip\noindent\begin{imageonly}\icon{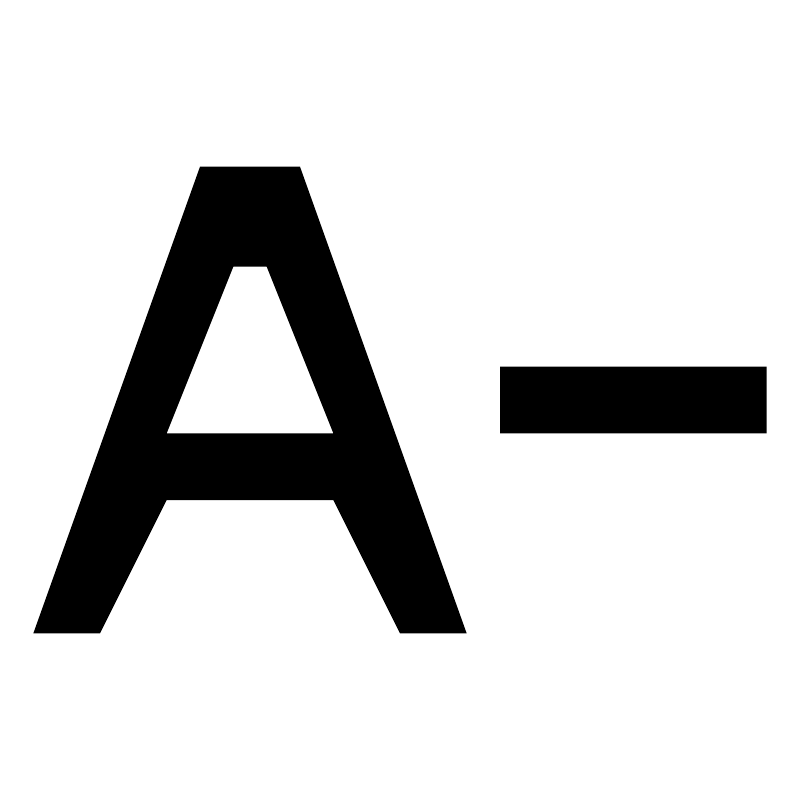}\end{imageonly} \textit{Singularizer and} \begin{imageonly}\icon{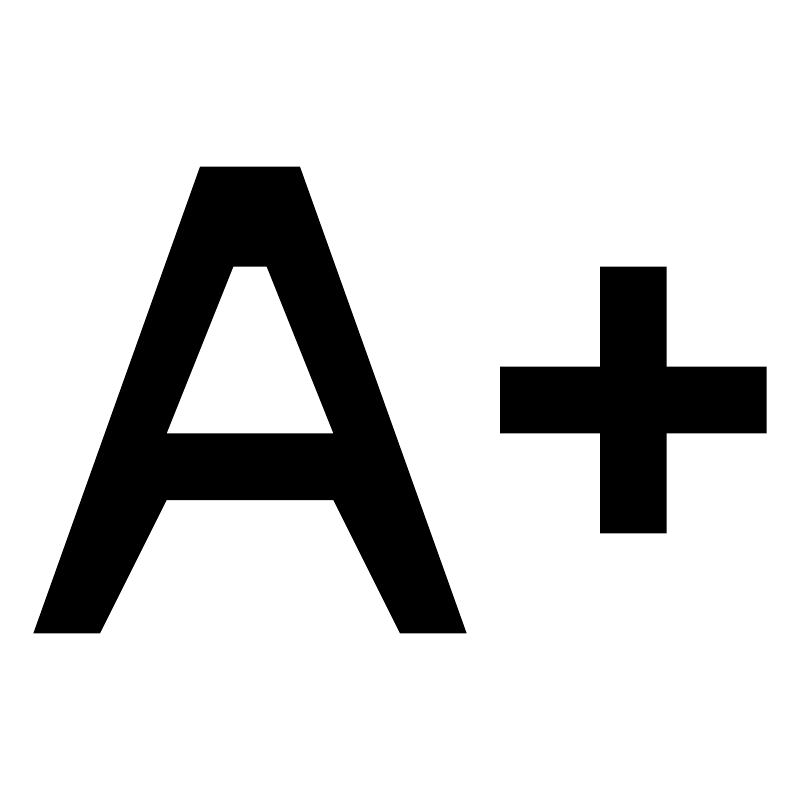}\end{imageonly} \textit{Pluralizer --} Singularizes or pluralizes the selection and updates the rest of the text to fix number agreements. For example, pluralizing ``She'' in the sentence ``She was tired of sitting on her own'' results in ``\textit{They were} tired of sitting on \textit{their} own'' where the verb and pronouns are also pluralized.

\medskip\noindent\raisebox{-2pt}{\includegraphics[width=9pt]{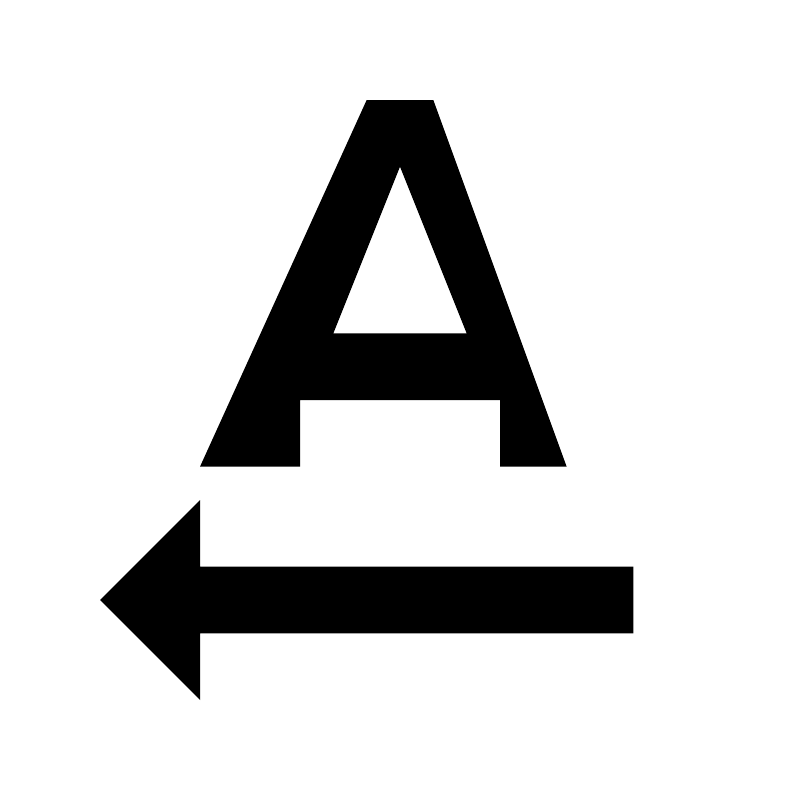}} \textit{Past,} \raisebox{-2pt}{\includegraphics[width=9pt]{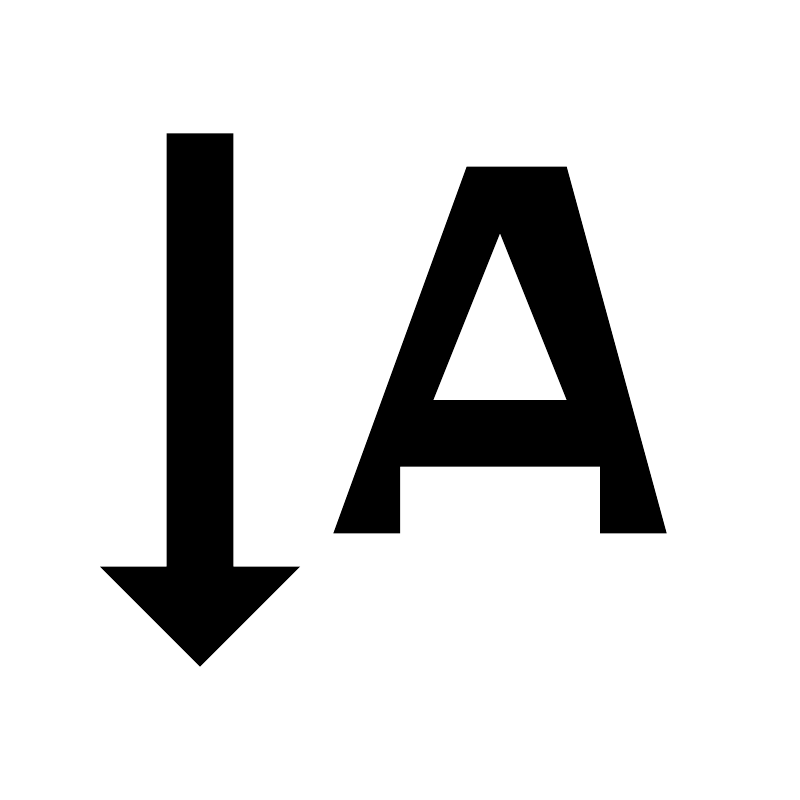}} \textit{Present, and} \begin{imageonly}\icon{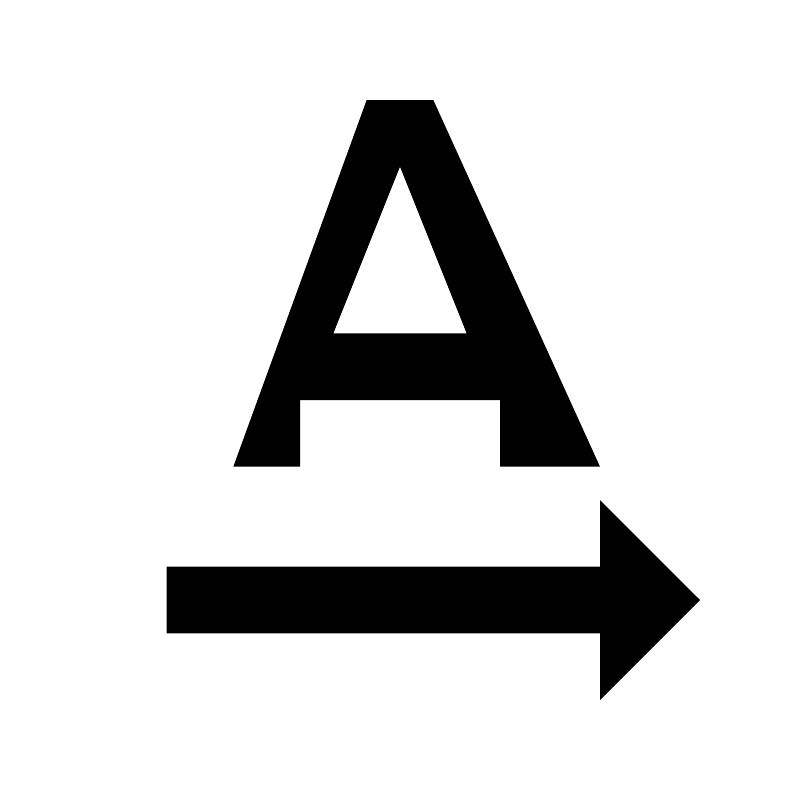}\end{imageonly}\textit{ Future Tense Changer --} Changes the tense of the selection and updates the rest of the text to fix tense agreements. For example, applying the future tense to ``was'' in  ``Alice was beginning to tire'' results in ``Alice \textit{will begin} to tire''.

\medskip\noindent\begin{imageonly}\icon{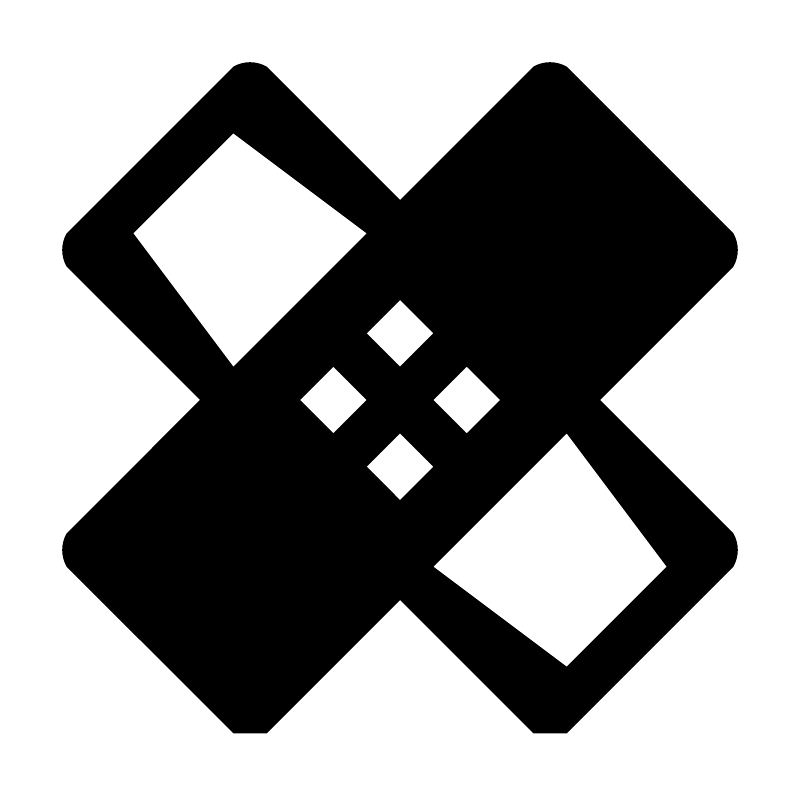}\end{imageonly} \textit{Repair --} Fixes the grammar, punctuation, spelling, and add necessary words to form valid sentences. For example, repairing ``Alice vry tire'' results in ``Alice is very tired''.

\medskip\noindent\begin{imageonly}\icon{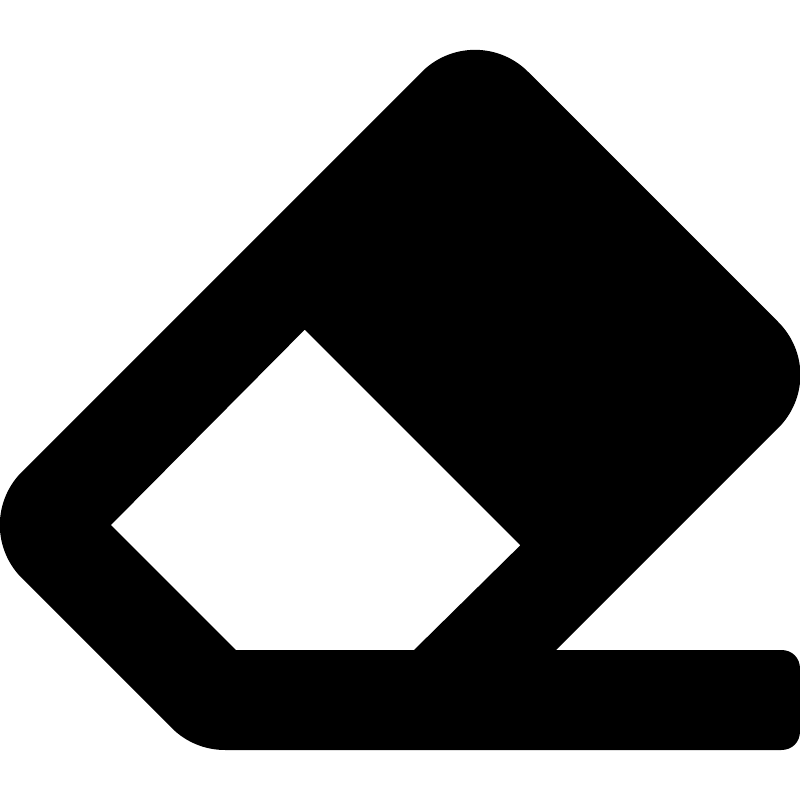}\end{imageonly} \textit{Eraser --} Removes the selected text and updates the rest of the text accordingly. Erasing makes sure the resulting text remains grammatically correct. Thus, it might restructure the sentence and slightly modify other parts. For example, erasing ``beginning'' in the sentence ``Alice was beginning to get very tired'' results in ``Alice was getting very tired''.

\medskip\noindent\begin{imageonly}\icon{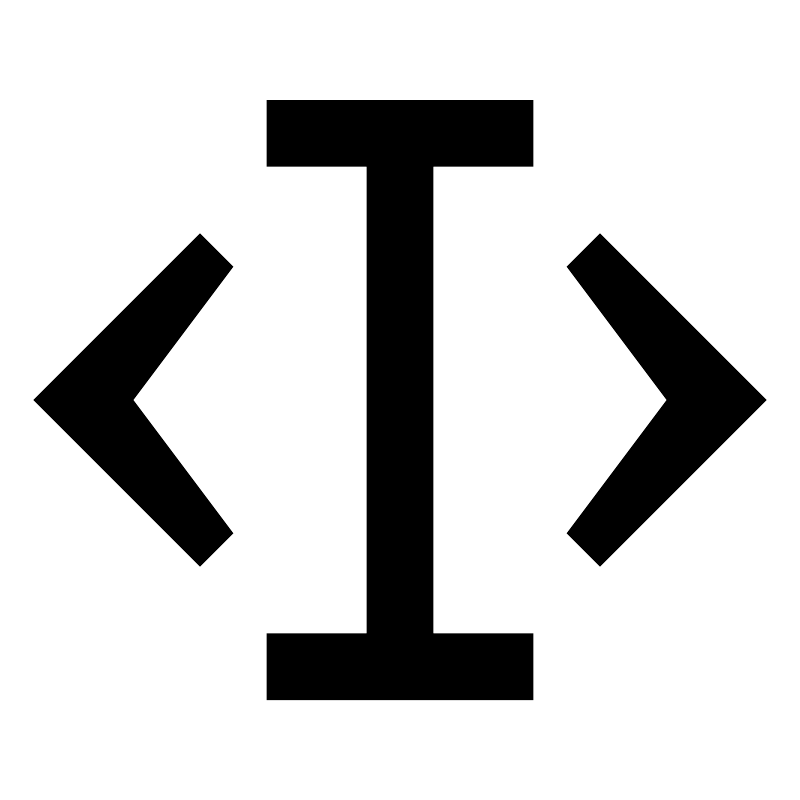}\end{imageonly} \textit{Prompt --} Applies a prompt on the selected text. After a text selection, a text input box opens to allow the prompt to be entered. A large language model executes the prompt and the result is immediately integrated into the text.

\subsection{Tone Picker}\label{sec:tonePicker}
\begin{figure}[htb]
    \includegraphics[width=.5\textwidth]{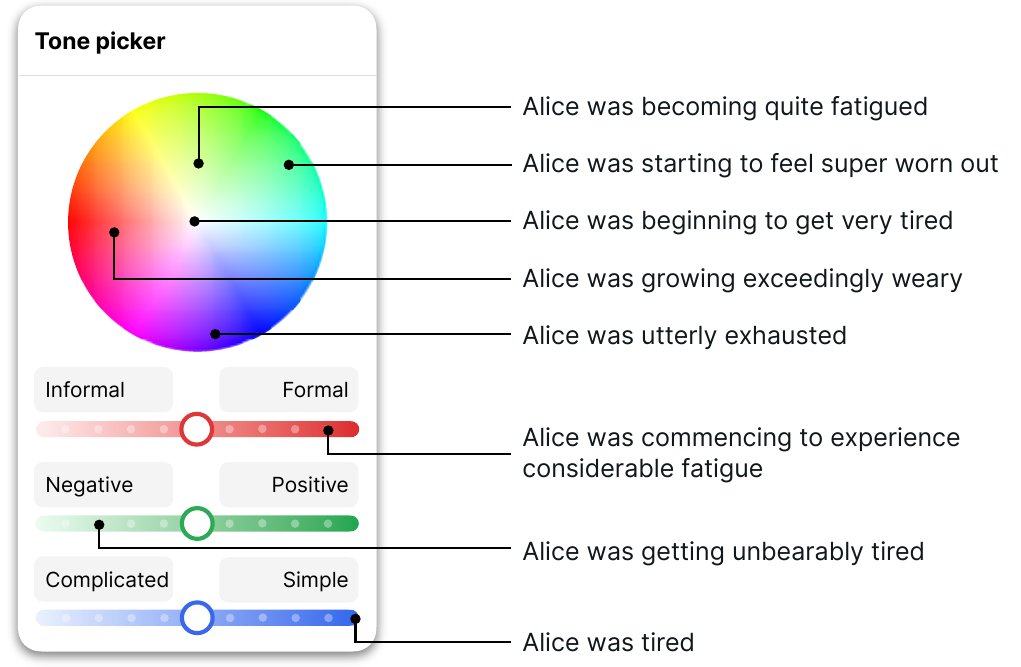}
    \caption{The tone picker allows exploring tone nuances of a sentence. Here, the selected sentence is ``Alice was beginning to get very tired'' and variations are explored by moving the thumb in the tone wheel and by changing the tone sliders.}
    \label{fig:tonePicker}
    \Description{Screenshot of the tone picker with arrows indicating what the text would become had the thumb been moved to that position in the space.}
\end{figure}

\name adapts traditional colour pickers to allow navigating the space of tones and support meaning-preserving permutations. Essentially, tones are mapped to colours, allowing traditional colour space representations to visually represent the space of tones. 
\name implements the RGB colour space.
A tone in \name is defined along three dimensions (\Cref{fig:tonePicker}). 
We choose ``formality'', ``sentiment'', and ``complexity'' and map them to red, green, and blue based on word-colour associations~\cite{kimLexichromeTextConstruction2020}. 
Blending a mix of colours results in the corresponding tone, for instance, a "purple" tone means formal (red) and simple (blue), and mostly negative (no green), see \Cref{fig:tonePicker}.
These tones were selected as defaults because of their orthogonality and their use in previous work on manipulating tones~\cite{jinDeepLearningText2021, okosoImpactToneAwareExplanations2024, huangCharacterizingSimilaritiesDivergences2024}.
These dimensions can be edited to explore other tone spaces, if needed. Further, we note that depending on user preferences and tasks, other colour spaces could be more suited.

\medskip\noindent\textit{Tone Sliders --} Changes the tone along one specific tonality dimension. The position of the thumb in the tone wheel is updated accordingly. Each of the three sliders ranges from 0 to 10, allowing to represent $11^3=1331$ different tones.

\medskip\noindent\textit{Tone Wheel --} Changes the tone along all three dimensions through one dragging motion in the colour wheel. Tone sliders are updated accordingly.
To help users understand the tone wheel, arrows are overlayed, indicating the direction of the strongest change for each of the three tonality dimensions.

\medskip\noindent\textit{Exploring Other Tones --} Tones in the colour picker are input text boxes that can be edited. Users can change the tones to explore new tone spaces, such as ``humour'' or ``Shakespearean'', or extend a scale to have finer granularity, such as going from ``formal'' to ``very formal'' to offer 10 levels of formality.

\medskip\noindent\textit{Changing the Tone After or Before the Selection --}
The tone picker supports both noun-verb and verb-noun command constructions depending on the order of actions~\cite[ch 3-3]{raskinHumaneInterfaceNew2000}. In a noun-verb construction, the text is selected first (noun), and following changes to the tone (verb) immediately update the passage, to rapidly explore different tone variations. In a verb-noun construction, the tone is selected beforehand (verb) and applied to future text selections (nouns) using the tone brush, to apply the same tone rapidly.

\subsection{Boolean Operations on Sentences}
\begin{figure*}[h]
    \includegraphics[width=\textwidth]{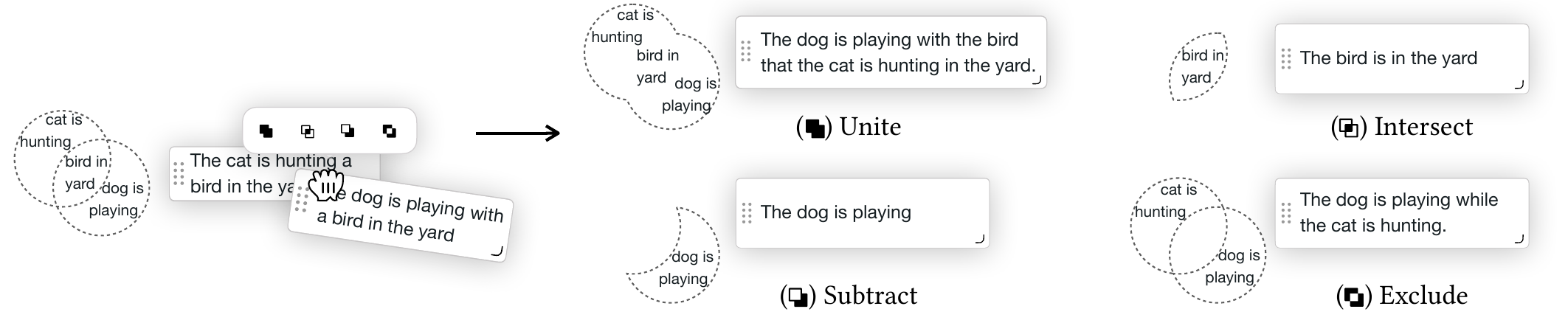}
    \caption[]{Dragging text over a passage allows merging the two passages in different ways: ({\raisebox{-2pt}{\includegraphics[width=9pt]{figures/tools/textfields_unite}}}) ``Unite'' merges the two passages; ({\raisebox{-2pt}{\includegraphics[width=9pt]{figures/tools/textfields_intersect}}}) ``Intersect'' preserves only the information shared by both passages; ({\raisebox{-2pt}{\includegraphics[width=9pt]{figures/tools/textfields_subtract}}}) ``Subtract'' removes the information in one passage from the other; and ({\raisebox{-2pt}{\includegraphics[width=9pt]{figures/tools/textfields_exclude}}}) ``Exclude'' removes the information shared by both passages.}
    \label{fig:booleanOperations}
    \Description{Screenshots with an abstract representation of what boolean operations do. The example tries to merge "The dog is playing with a bid in the yard" with "The cat is hunting a bid in the yard" which results in "The dog is playing with the bird that the cat is hunting in the yard" when uniting, "The dog is playing" when subtracting, "The bird is in the yard" when intersecting, and "the dog is playing while the cat is hunting" when excluding.}
\end{figure*}

Overlapping text can be merged in different ways to support micro and macro text changes. These merges are based on the ideas expressed in the text. \name supports the following five methods to merge two pieces of text directly inspired by boolean operations on shapes (\Cref{fig:booleanOperations}).

\medskip\noindent\begin{imageonly}\icon{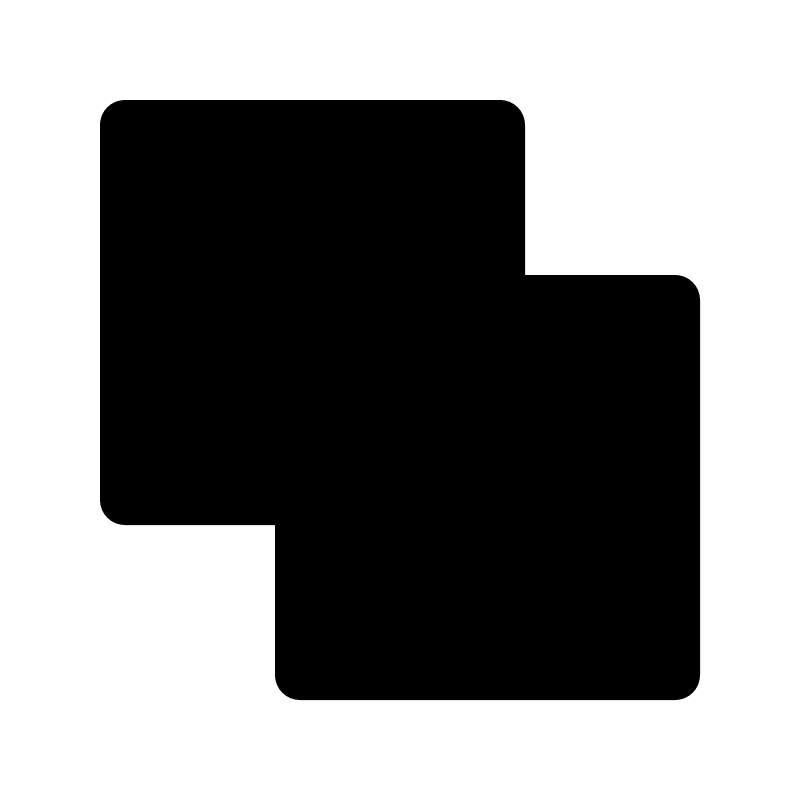}\end{imageonly} \textit{Unite --} Merges the two passages so that the resulting text is grammatically correct and includes the ideas from both passages. For example, merging  ``the dog is playing'' with ``the cat is playing'' results in ``the dog and the cat are playing''.

\medskip\noindent\begin{imageonly}\icon{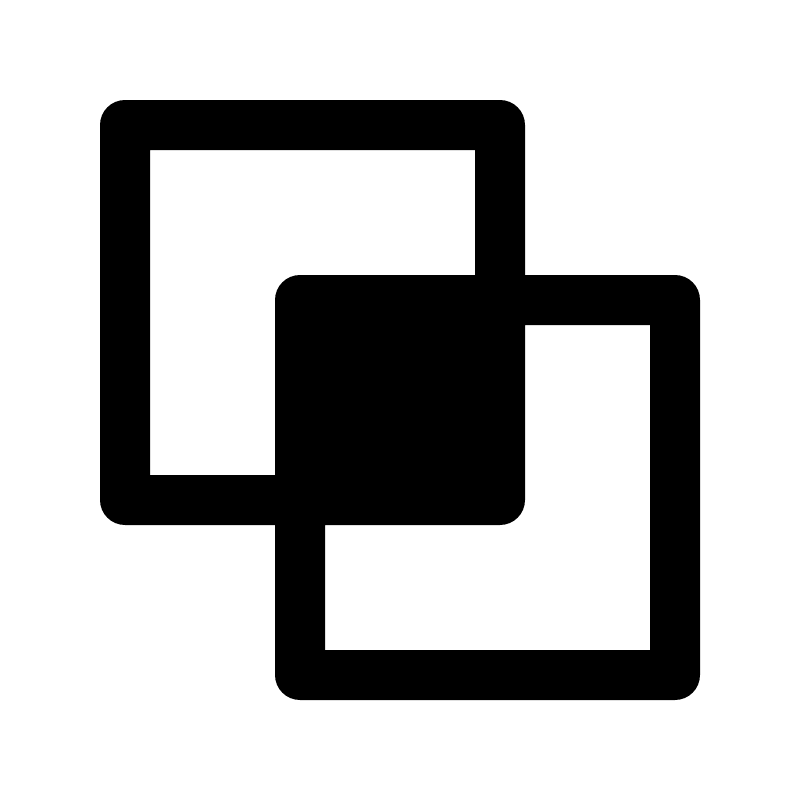}\end{imageonly} \textit{Intersect --} Forms a grammatically correct text that expresses the ideas shared by both text passages. For example, intersecting ``the cat and the dog are playing in the yard'' and ``the dog is chasing the cat in the yard'' results in ``the dog is in the yard''.

\medskip\noindent\begin{imageonly}\icon{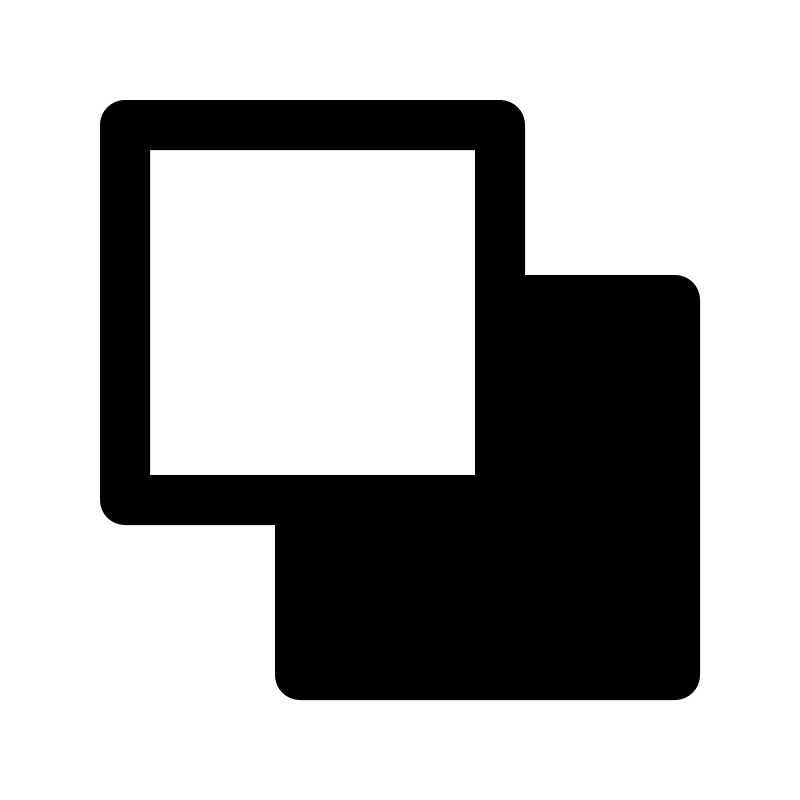}\end{imageonly} \textit{Subtract --} Removes the ideas expressed in the first passage from the second passage. For example, subtracting ``it's raining today'' from ``she wears her coat because of the wet weather'' results in ``she wears her coat''.

\medskip\noindent\begin{imageonly}\icon{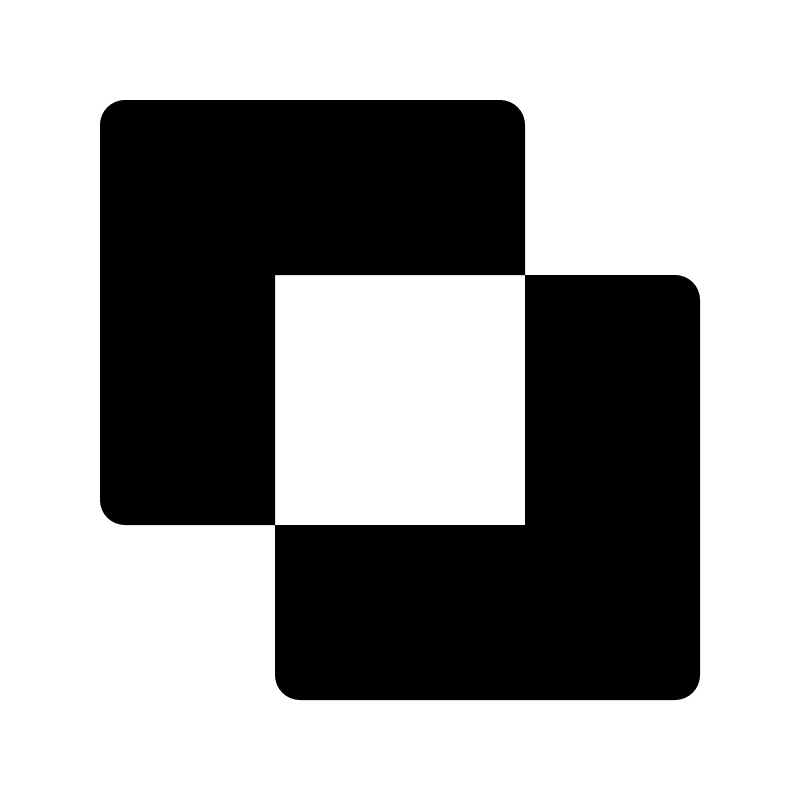}\end{imageonly} \textit{Exclude --} Removes the ideas that are shared in both passages to preserve only the non-redundant ideas. For example, excluding ``Tom likes both chocolate and vanilla ice cream'' from ``Tom enjoys chocolate ice cream'' results in ``Tom likes vanilla ice cream''.

\medskip\noindent\begin{imageonly}\icon{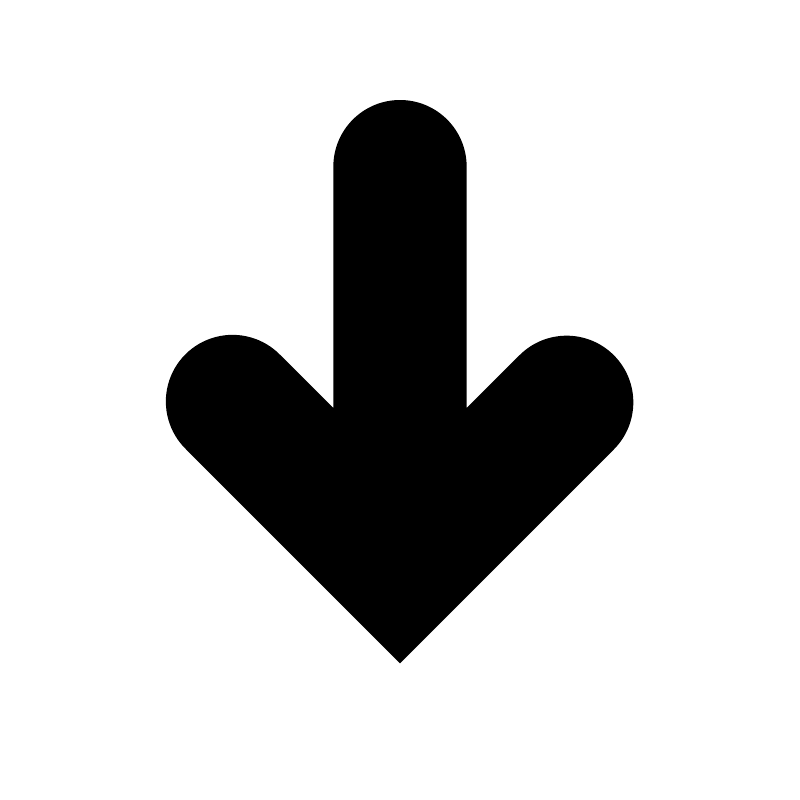}\end{imageonly} \textit{Insert --} Naively inserts the text passage at the exact caret position where the text was dropped, similar to pasting the text in a regular word processor. This allows converting a freely moving text selection back into text on the page.

\subsection{Text Layers}
Layers were originally introduced to drawing software to support non-destructive editing and allow picture parts to be edited independently~\cite{andrewsAdobePhotoshopElements2006}. These layers were inspired from the transparent plastic sheets used by the animation industry to compose scenes. But they rapidly became more powerful~\cite{jacksonCompositingDigitalIllustration2015} and even essential to support the organization and versioning of drawings~\cite{tsandilasBricoSketchMixingPaper2015}. Writers share similar needs in that they want more organizational support~\cite{Scrivener} and non-destructive edits that can be retrieved and reversed~\cite{rezaABScribeRapidExploration2024}.

\begin{figure}[hbt]
    \includegraphics[width=\columnwidth]{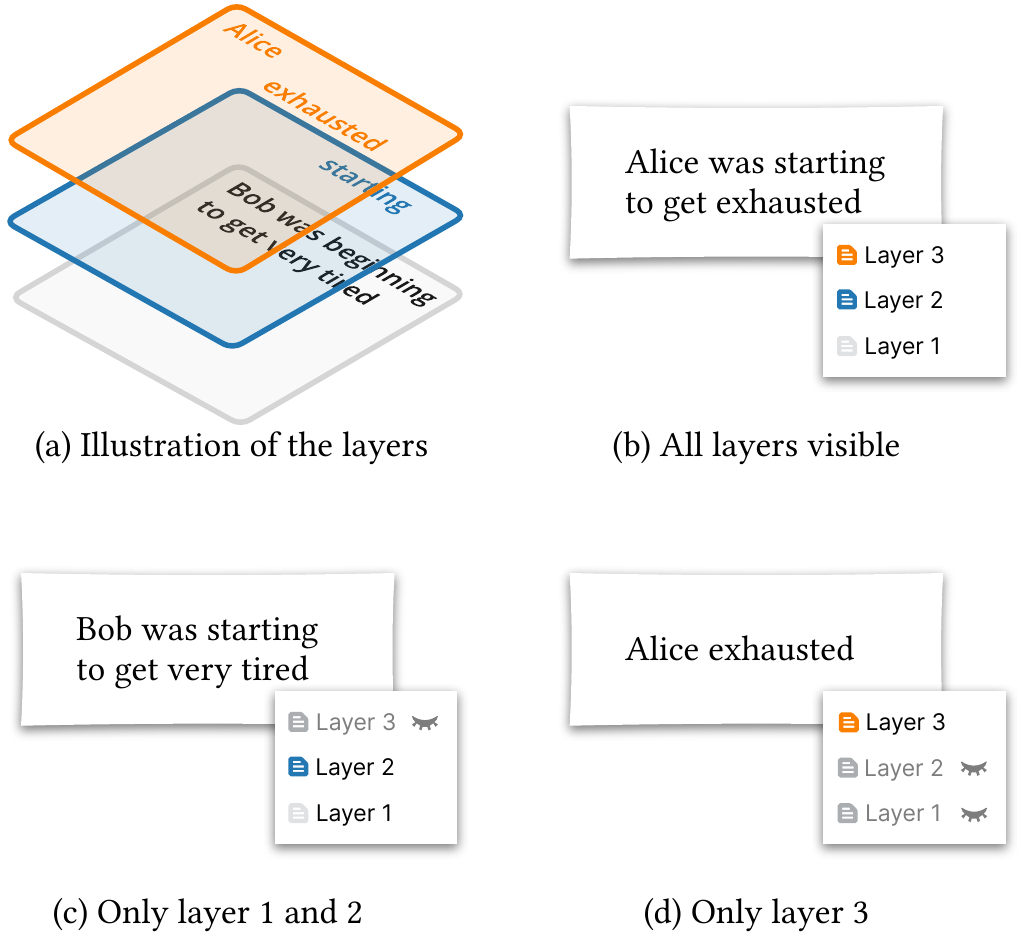}
    \caption{Text layers are stacked on top of each other: (a) each layer may store text anchored at different text locations; (b) layers at the top might hide the text underneath; (c) layers can be hidden; and (d) text is reflowed to avoid gaps}
    \label{fig:layers}
    \Description{Illustration of the layers. The layers are demonstrated through a 3D visualization of the stack of transparent layers. Depending on which layer is visible, the text changes.}
\end{figure}

In \name, we propose to rely on the familiar layer metaphor.
Text from documents can be organized in layers. These layers are similar to layers from drawing software in that they can be viewed as transparent sheets stacked on top of each other. However, unlike drawing layers, text layers store modifications that are anchored in the text from layers underneath it (\Cref{fig:layers}). 
The reason is that, unlike images, text is positioned relative to other text (e.g., a sentence is positioned after another sentence, but it makes little sense for it to be placed at specific coordinates). This means that the position of the text in a layer is relative to the text of the preceding layers. A modification to any of the preceding layers might reflow the text of the following layers. And layers can hide the text below them.

\medskip\noindent\begin{imageonly}\icon{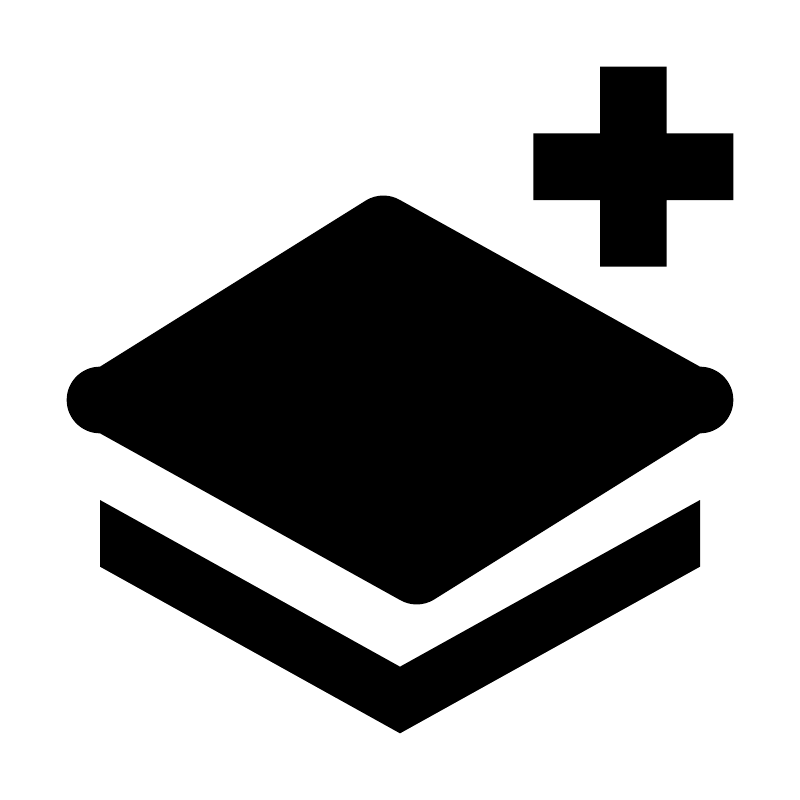}\end{imageonly} \textit{Adding a Layer --} Adds an empty layer at the top of the stack. Then, if this layer is selected, all new text modifications are stored in that layer. For example, text from the preceding layers can be modified or removed, and new text can be added, in which case it will be anchored within the text of the preceding layers.

\medskip\noindent\begin{imageonly}\icon{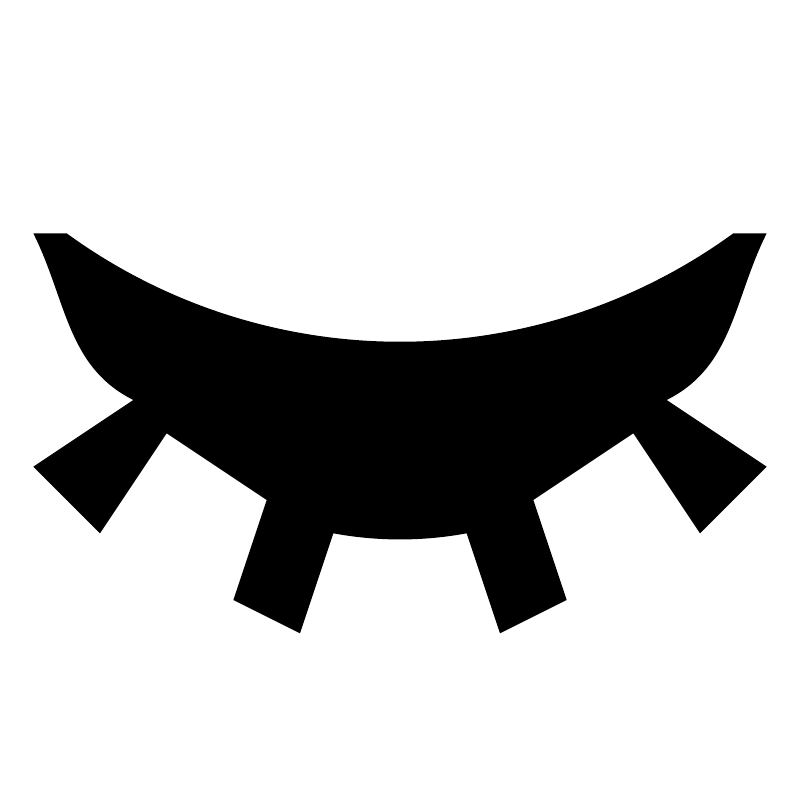}\end{imageonly} \textit{Hiding and }\begin{imageonly}\icon{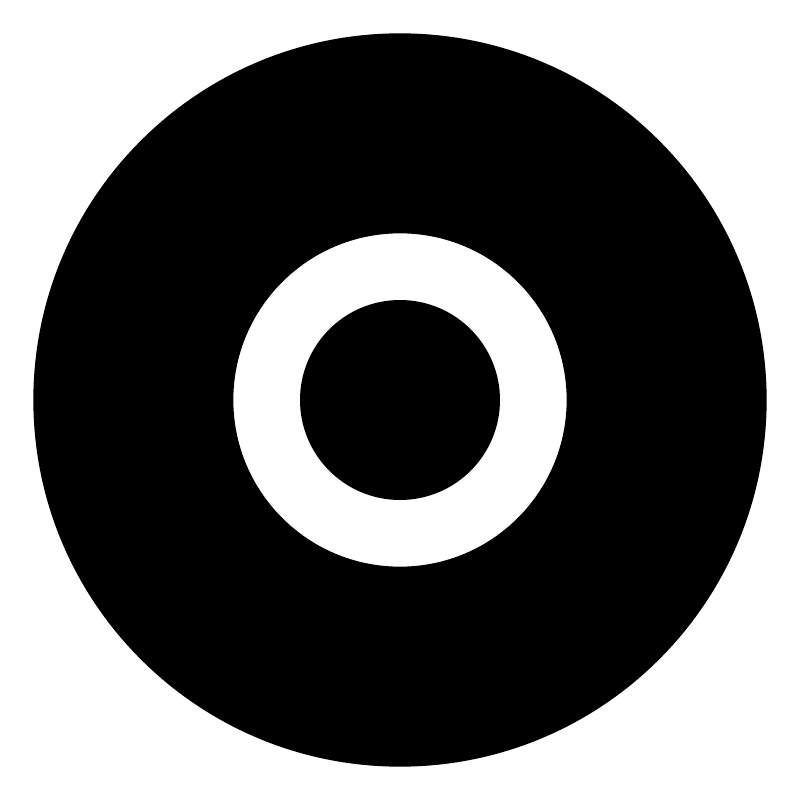}\end{imageonly} \textit{Showing a Layer --} Hides/shows the modifications stored in a layer. For example, if a layer modified the text, hiding this layer restores the original text. And if a layer added some text, hiding the layer removes the added text.

\medskip\noindent\textit{Reordering Layers --} Changes the order in which layers are shown. Reordering is done by a drag-and-drop interaction where the final position of the layer is decided by where it is dropped in the list. The order of layers decides what is visible. For example, the top layer (i.e., the first in the list) is always visible because it appears above all other layers.

\subsection{Implementation}
The \name interface emulates the look and feel of traditional drawing software, except the main canvas is replaced by a text input field (\Cref{fig:teaser}). Under the hood, it uses a large language model as well as more traditional natural language processing techniques. 

\name is implemented in TypeScript using React~\cite{ReactJavaScriptLibrary2013} for the interface, Slate.js~\cite{SlateJS} for the text editor, NextUI~\cite{NextUI} for the graphical components, and the OpenAI library~\cite{openaiOpenAINodeAPI2023} to execute prompts using OpenAI's ``gpt-4o'' model. A live demo and the source code are available\footnote{Live demo and code: \url{https://github.com/m-damien/Textoshop}}.

\subsubsection{Highlighting Text Changes} To help identify text changes, we added an animation to highlight character additions and deletions. The animation follows the guidelines from the transitioning technique between text revisions proposed by \citet{chevalierUsingTextAnimated2010}: deleted characters appear first, in red, and progressively shrink and fade out. Added characters appear second, in green, and progressively grow and fade in until matching the size of the rest of the text. The whole animation lasts about one second.

\subsubsection{Preserving Semantic Meaning} Certain modifications, such as resizing or relocating text, might result in text that is grammatically correct but with a different semantical meaning. We implemented several mechanisms to mitigate the loss of semantic meaning. For example, when resizing, the original sentence is saved and always used for subsequent prompts sent to the model. This means that even if the sentence is reduced to a couple of words, expanding it again will be based on the original longer sentence. Similarly, when relocating a passage, the model is prompted with the sentences surrounding the final location so that the sentence is adjusted to blend into its new location.

\subsubsection{Prompting Strategies and Heuristics} Our intent with the implementation was to have robust, fast, and reliable tools to allow testing the concept behind \name. While the implementation could perhaps be improved, we currently rely on engineered prompts and heuristics. 
For example, tools operating on a selection were optimized to only send the selection to the model and reintegrate the results by making it match the formatting of the original text. Another example is the resizing tool, which first splits the selection into sentences and prompts the model to shorten or expand each sentence in parallel. In our tests, GPT-4o was unreliable when prompted to resize a text by a certain amount of words. To boost the reliability of the resizing, the model was prompted to resize the same text eight times using a slightly different number of words. Then, the system calculated all possible combinations of the resized parts and selected the one closest to the target length.

\section{User Study}
We ran a two-part user study to learn about the usability and usefulness of our approach. Specifically, we tried to answer the following questions: How do people use the features of \name to accomplish typical writing tasks? How does their knowledge of drawing software help them understand \name? Is \name easier to use than existing solutions?

The first part of the study allowed us to isolate the usability of different features in a controlled setting. Each participant accomplished a series of tasks using either \name or a baseline interface. For the baseline, given that no existing solutions are directly comparable to \name in terms of features, we opted for an interface composed of a traditional text editor and an integrated replica of the ChatGPT interface, which participants could supplement with any external software as they saw fit. 

The second part of the study was more exploratory to observe how participants use the tool as a whole in a more holistic and unguided way. Participants used the fully-featured version of \name on their own text with no other instruction than to explore the tool and its features to accomplish their text editing goals.

\subsection{Participants}
We recruited 12 participants from our local community (age range: 23 to 35, Mdn=26; 7 self-identified as female and 5 as male). On a 5-point scale from 1-never to 5-very frequently, participants reported how often they engaged in writing activities (Mdn=5, SD=.7), how often they use generative AI to help them during writing activities (Mdn=2.5, SD=1.5), and how often they use drawing software  (Mdn=3, SD=1). Most of their writing was academic (12), professional (11), and creative (2). On a 5-point scale from 1-not at all familiar to 5-extremely familiar, they also reported their familiarity with specific drawing software features such as colour pickers (Mdn=5, SD=.8), drawing tools (Mdn=4, SD=.7), layers (Mdn=4, SD=1.1), and boolean operations on shapes (Mdn=4, SD=.6).

\subsection{Apparatus}
We ran the study remotely using the Zoom conference software. Participants used their personal computers and used a modern web browser to open the experiment website. The whole session was audio and screen recorded, and the study software logged all interactions with the system (e.g., clicks, features used, answers to questions). Sessions were about an hour long. Participants received a CAD\$20 gift card in appreciation for their time. Our institutional ethics board approved the study.

\subsection{Procedure}
Each participant used two \textsc{interface}s (\name and baseline). For each \textsc{interface}, the participant completed 4 \textsc{tasks} (shortening and expanding a story, organizing and integrating text fragments, changing the tone of dialogue, and using a template to write emails).

\medskip\noindent\textit{Introduction (\textasciitilde{}2min) --} The participant completed the consent form and demographics questionnaire and was told they would use two different interfaces to accomplish a series of text editing tasks.

\medskip\noindent\textit{Video Tutorial} (2 \textsc{interface}s $\times$ 4 \textsc{task}s $\times$ \textit{20sec) --} Before each task, the participant watched a 20-second video demonstrating the feature they are about to use. The videos demonstrate features on abstract tasks, akin to the typical tool previews found in modern software~\cite{grossmanToolClipsInvestigationContextual2010}. The video for the baseline condition demonstrates the use of our replica of ChatGPT, and the participant was allowed to skip it after having watched it once. No video was provided for the external software the participant might have chosen to use.

\medskip\noindent\textit{Text Editing Tasks} (2 \textsc{interface}s $\times$ 4 \textsc{task}s $\times$ \textit{4min) --} After watching the video mini-tutorial, the participant was asked to accomplish a text editing task using a given \textsc{interface}. In the baseline condition, besides the provided integrated version of ChatGPT, participants were encouraged to use external software of their choosing to accomplish the task if they felt it would be easier. Instructions for the task were shown in the bottom left corner and consisted of a short textual description. The participant had to manually indicate when done with the task by pressing a button. Otherwise, after 4 minutes of working on the task, the participant was asked to stop and move to the next task. The participant then had to indicate how successful they felt in accomplishing the task on a 5-point semantic differential scale from 1-unsuccessful to 5-successful.

\medskip\noindent\textit{Questionnaire About Interface Used} (2 \textsc{interface}s $\times$ \textit{2min) --} After completing the 4 tasks with an \textsc{interface}, the participant completed a system usability scale (SUS) questionnaire~\cite{brookeSUSQuickDirty1995} and answered 5-point scale questions related to the difficulty of the tasks they just completed. The second and final questionnaire also included comparative scale questions asking the participant to report the interface they preferred to accomplish each task.

\medskip\noindent\textit{Free-form Exploration with \name} (\textasciitilde{}\textit{10min) --} In a second part, the participant was invited to use the fully-featured version of \name on their own text. Before the study, the participant had been instructed to bring a text they ``would be interested in editing (about 4-5 paragraphs long)''. In case they did not bring text, they were also allowed to choose amongst a collection including Wikipedia articles and passages from novels. Because we meant for this part to be a free-form exploration, the participant received no instruction other than ``to explore the tool and its features to accomplish their editing goals''. Given that some of the features of \name had not been used nor demonstrated in the previous part, the participant could also ask for an explanation of these features.

\medskip\noindent\textit{Semi-structured Interview} (\textasciitilde{}\textit{10min) --} After finishing the free-form exploration, the participant engaged in a semi-structured interview with the experimenter. Questions were open-ended and related to their experience with the system, their preferences, and their impressions on the design of the tool and its usability.

\subsection{Design and Tasks}
The study followed a within-subject design. Interface order was counterbalanced. To avoid participants doing the same tasks with both interfaces, we designed two variations of each task using different but comparable texts. The order of the tasks followed a balanced Latin square, and the variation used was picked at random. 

We designed four tasks such that they cover the issues with traditional ways of editing text (\cref{sec:issues}) and involve as many of the features of \name as possible. The tasks were designed to be realistic, require multiple steps, and ask of participants to express higher-level intents rather than simple editing operations. Rather than evaluating all tools, our main question was whether participants understand and successfully use features building on the drawing software analogy. The four tasks were as follows.

\medskip\noindent\textit{Shortening/Expanding a Summary --} Given a summary of a well-known novel, the participant has to shorten and expand specific sentences and paragraphs. For example, they need to compress a paragraph to avoid the short last line (``orphan'') while preserving the meaning. Or they need to expand upon a specific sentence in another paragraph so that it covers two lines and fills the whole width of the page. With \name, the participant can use direct manipulation to resize text selections. 
This task emulates the frequent need for writers to change the length of their text so that it fits within a spatial layout (e.g., posters and presentations) or remains below a character or page limit.

\medskip\noindent\textit{Changing the Tone of Dialogues --} 
Given a dialogue between two fictional characters, the participant has to change the tone of the dialogue of one of these characters (e.g., make their dialogues more informal and positive) and then explore at least four variations of the other character's last dialogue.
With \name, the participant can use the tone picker and tone-brush. 
This task was to test the tone picker metaphor. It emulates copyediting situations when the tone of a document has to be uniformized (e.g., integrating changes from different writers) and when the tone has to be explored precisely to match a certain intent (e.g., writing an email to different audiences).

\medskip\noindent\textit{Organizing and Integrating Text Fragments --} Given a simple story and a bullet list with 6 text fragments, the participant has to sort the text fragments depending on the character they are related to and then integrate these fragments into the story. Some text fragments have to be merged into specific parts of the text, while others are concepts to remove from the story.  In the \name condition, participants can arrange the text fragments in and off the page and can also use boolean operations to unite, intersect, subtract, or exclude text fragments. This task was to test the metaphor of the canvas and floating elements, as well as boolean operations on text fragments. It emulates editing the flow of ideas and rapidly iterating on different ways to structure these ideas.

\medskip\noindent\textit{Using an Email Template to Answer Customers --} 
Given the profile of two customers as well as the email template used by a fictional company to answer their customers, the participant has to generate an email in response to each customer. The template includes pre-written paragraphs, which should be selected if they apply to that specific customer, and has to be edited further to include the name of the customer and some specific information, such as amounts and durations. With \name, layers are already created with the text of the email template; to form the final emails, the participant has to show/hide different layers and apply minor edits to the layers. With the baseline interface, the email template lists all the possible phrasings to answer a customer organized in sections; to form the final emails, the participant has to copy and paste them in order. This task was to test the layer metaphor and participants' mental model of layers when editing. It simulates having to produce different text versions simultaneously based on different audiences, such as email communication or educational resources.

\subsection{Data Analysis}
Considering ongoing debates about statistics in HCI~\cite{dragicevicFairStatisticalCommunication2016, groupTransparentStatisticsGuidelines2019}, we report both confidence intervals (CI) and p-values. However, we base our conclusions on the 95\% confidence intervals (CI) on mean differences, which provide information about both the presence and size of effects. All confidence intervals are calculated using the studentized bootstrapping method because it is the most robust for within-subject study designs of about 12 participants~\cite{massonStatslatorInteractiveTranslation2023, zhuAssessingComparingAccuracy2018}. All p-values are calculated using a Wilcoxon signed-rank test because it is less likely to yield false positives compared to parametric alternatives when dealing with data whose distribution is unknown and possibly heavily skewed~\cite{bridgeIncreasingPhysiciansAwareness1999}. The bootstrapped CIs are calculated in Python using arch 5.5.0~\cite{kevin_sheppard_2023_7975104} and using 10,000 replications. The Wilcoxon signed-rank tests are calculated using scipy 1.10.1~\cite{scipy} with ties being discarded.

\newcommand{\stat}[1]{{\small#1}}

\subsection{Results from the Text Editing Tasks}
\begin{figure*}[t]
    \includegraphics[width=\textwidth]{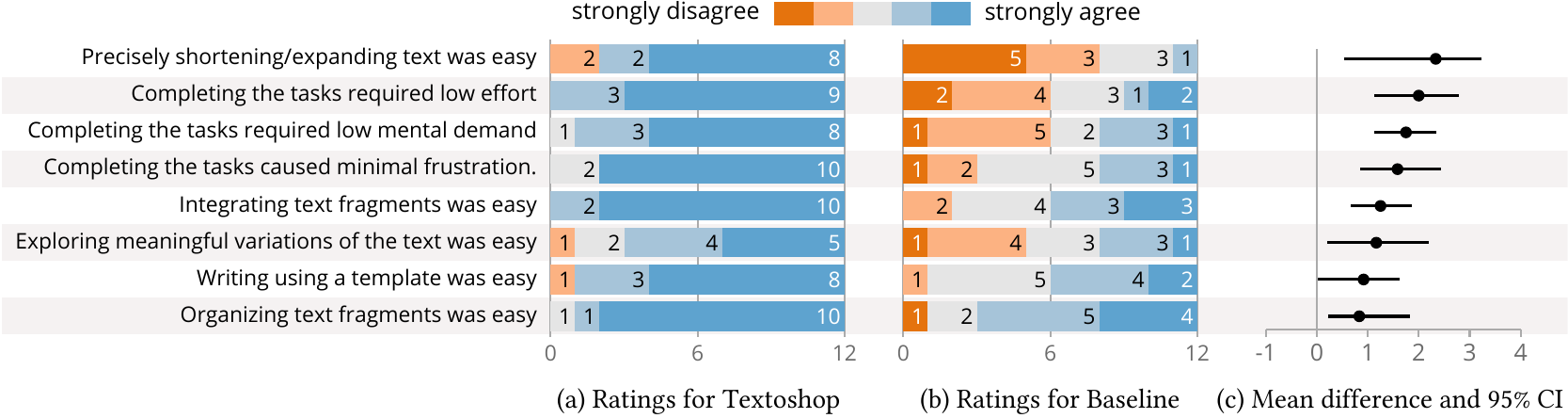}
    \caption{Participants' response when rating the 5-point statements for (a) \name and (b) baseline. (c) Dots are the mean differences of \name compared to baseline. Bars are the 95\% CIs calculated with the studentized bootstrap method.}
    \label{fig:statements}
    \Description{Stacked bar charts of the responses from participants to the 5-point scale statements. The majority of the ratings for Textoshop are in the agree and strongly agree zone, whereas the ratings for the baseline are more split, with some questions such as "Precisely shortening/expanding text was easy" being mostly rated strongly disagree and disagree.}
\end{figure*}
Participants completed the tasks without major difficulties. 
During the baseline condition, and despite being encouraged to use external tools, only P11 used Grammarly. Other participants used a mix of the provided replica of ChatGPT and manual editing.

\subsubsection{Success rate} On a 5-point semantic differential scale from 1-unsuccessful to 5-successful, participants rated being more successful at accomplishing the tasks using \name than baseline by 0.71 \stat{(95\% CI [0.27, 1.17], M=4.6 vs M=3.9, p=.015)}.

\subsubsection{Usability} On the system usability scale and for the given tasks, \name received a 90 (typically considered ``excellent''~\cite{bangorEmpiricalEvaluationSystem2008}) while the baseline received a 69 (typically considered ``ok''). This corresponds to a difference of 21.25 \stat{(95\% CI [12.5, 32.91], p<0.001)}.

\subsubsection{Time and Prompts Used} Although productivity was not our focus, we provide indications of completion time and prompts used as a loose indicator of the tedium involved in completing a task. Participants completed tasks faster using \name than baseline by \prettytimestamp{55} \stat{(95\% CI [\prettytimestamp{26}, \prettytimestamp{86}], M=\prettytimestamp{143} vs M=\prettytimestamp{199}, p=0.003)}. Ranked from largest to smallest effects:
``Shortening/expanding a summary'' by \prettytimestamp{89}, 
``Using an email template'' by \prettytimestamp{79},
``Changing the tone of dialogues'' by \prettytimestamp{41}, and
``Organizing and integrating text fragments'' by \prettytimestamp{13}. Additionally, in the baseline condition, participants used 2.7 prompts per task on average, and these prompts were about 269 characters long. Considering participants did not use prompts in the \name condition, we discuss participants' behaviours in \cref{sec:observations}.

\subsubsection{Questionnaires} Participants responded to all eight 5-point statements in favour of \name on average. \Cref{fig:statements} shows the breakdown of participants' answers. 
Ranked from largest to smallest effects: 
``Precisely shortening/expanding text was easy'' by 2.33 \stat{(95\% CI [0.54, 3.23], p=.007)}, 
``Completing the tasks required low effort'' by 2.0 \stat{(95\% CI [1.13, 2.79], p=.004)}, 
``Completing the tasks required low mental demand'' by 1.75 \stat{(95\% CI [1.13, 2.34], p=.003)}, 
``Completing the tasks caused minimal frustration.'' by 1.58 \stat{(95\% CI [0.85, 2.44], p=.007)}, 
``Integrating text fragments was easy'' by 1.25 \stat{(95\% CI [0.66, 1.86], p=.006)}, 
``Exploring meaningful variations of the text was easy'' by 1.17 \stat{(95\% CI [0.2, 2.2], p=.026)}, 
``Writing using a template was easy'' by 0.92 \stat{(95\% CI [-0.02, 1.62], p=.053)}, 
``Organizing text fragments was easy'' by .83 \stat{(95\% CI [0.21, 1.83], p=.033)}.

\subsubsection{Preferences} When asked about their preferred \textsc{interface} for each task, participants unanimously rated \name as  ``preferred'' or ``strongly preferred'' when ``integrating text fragments''. All participants also preferred \name when ``precisely shortening/expanding text'' except P10, who did not have a preference for either interface. For other tasks, while the majority preferred \name, some participants expressed other preferences: when ``writing using a template'' P8 preferred the baseline interface; when ``organizing text fragments'' P3 preferred and P8 strongly preferred the baseline; and when ``exploring meaningful variations of text'' P7, P8, and P12 preferred and P3 strongly preferred the baseline. We discuss the participants' reasons for these preferences in \Cref{sec:preferences}.

\subsection{Results from the Free-form Exploration}
\begin{figure*}[t]
    \includegraphics[width=\textwidth]{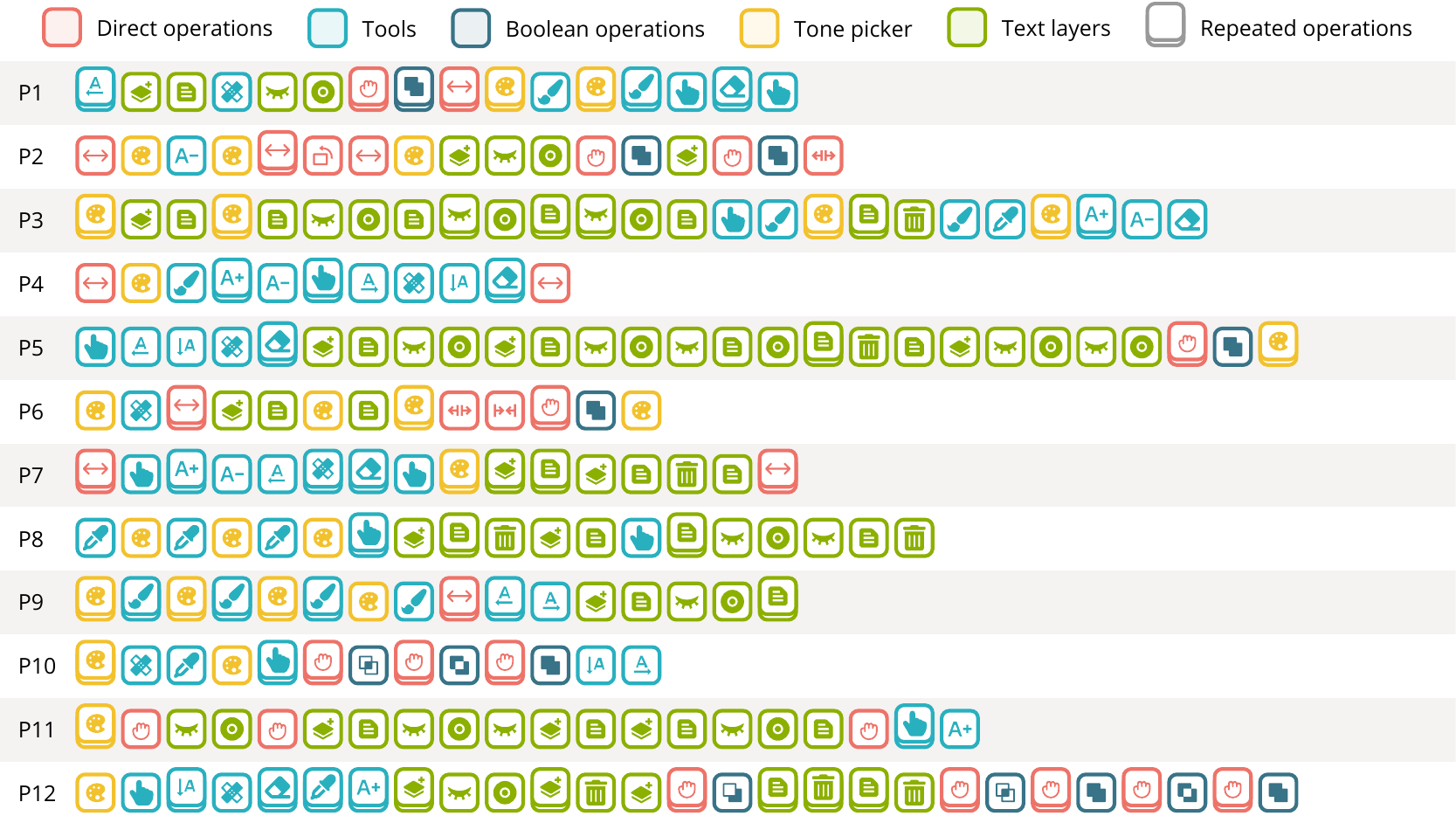}
    \caption{Sequence of operations done by participants using \name during the free-form exploration. Repeated sequences of the same feature are represented as a stacked pile. Refer to \cref{sec:textoshop} for the meaning of the icons.}
    \label{fig:operations}
    \Description{List of commands used by each participant during the free-form part of the study. Overall, participants covered all types of commands. Boolean operations were the least frequent, with P3, P4, P7, P8, P9, and P10 never using them. P4 and P10 are the only ones that did not use layers. And P8 was the only one that did not use the tone wheel. Direct manipulations were used by all at least once, if only to move or resize sentences.}
\end{figure*}

In the free-form exploration, participants chose texts related to creative story writing (P2, P4, P5, P9), informative writing (P3, P10, P11, P12), academic writing (P1, P6, P7), and personal writing (P8). 
\Cref{fig:operations} shows the sequence of features used by each participant.
Below, we report some of the strategies and behaviours observed.

\subsubsection{Iterating on a sentence by alternating tools} Participants used the tools to iterate on the wording of some specific sentences, much like they would iterate on the colours, texture, and organization of elements in drawing software. However, unlike visual artists who often apply colours and texture only after sketching~\cite{tsandilasBricoSketchMixingPaper2015}, with \name, tone exploration happened right from the beginning. Often, participants' workflow involved alternating between trying different tones, shortening the sentence, and smudging some parts to be paraphrased. The order of these steps was not always the same, but the tone picker was often used first, and then the other tools would allow editing with more precision. With the exception of P2, who frequently rotated and split the sentences, other tools were seldom used to rewrite a sentence.

\subsubsection{Experimenting with alternative organizations of ideas} P5, P10, and P12 reorganized the ideas multiple times by taking a passage from one paragraph and integrating it into a different paragraph or sentence using the boolean operations. P6 did the same but created their own text fragments to add an idea to an existing paragraph. Finally, P1 went one step further and deconstructed a whole paragraph into passages. Then, they organized these passages in the margin to try a different organization of the arguments and then reintegrated the passages into the main text using the ``insert'' and ``unite'' features. This suggests \name encouraged writers to work on the flow of ideas even after the writing is fully-fledged into sentences and paragraphs. This is in contrast with traditional writing, where restructuring mostly happens in the planning phase, when the draft is only an outline~\cite{flowerProblemSolvingStrategiesWriting1977}.

\subsubsection{Using layers to identify changes, store AI modifications, or add structure} Layers were used in multiple ways, most differing from how they are used in drawing software. For example, P1, P8, and P9 used one layer to store all the modifications made using the tools, allowing them to preserve the integrity of their original text and easily identify AI changes by switching the visibility of the layer. P3 used layers to store alternative versions of paragraphs (in this case, storing a more informal version). Perhaps the closest use of text layers to drawing layers that we have observed is to use them to structure the text. For example, P5, P11, and P12 created layers to store the different paragraphs, passages, or main ideas. P7 used layers to store the examples mentioned in the text. And P6 used a layer to store the statistics reported in the text so that they could read and edit their results section while hiding the means or p-values reported in the text (by hiding the statistics layer). This differs from traditional writing, where text cannot usually be teased apart in semantically meaningful ways. It is akin to the organization proposed by systems like Scrivener~\cite{Scrivener} except with a much finer granularity than just chapters and sections.

\subsubsection{Non-linear editing and reviewing} While it is known that reviewing and editing are non-linear processes~\cite{flowerProblemSolvingStrategiesWriting1977}, \name seemed to be more favourable to this behaviour than existing word processors. P3, P4, P6, P7, P9, P11, and P12 seemed to follow a tool-driven approach to editing, where selecting a new tool triggered a new pass on the document. For example, after attending to changes at the end of the document, P1 selected the eraser and went through the document from the beginning, this time trying to erase words that may not be essential, such as adverbs. Still, P2, P5, P8, and P10 proceeded more linearly, reading through the documents and switching between tools as needed, in a process akin to editing in a traditional word processor.

\section{Discussion}
We first answer our main research questions using the results and participants' comments. Then, we open the discussion to themes stemming from observations and discussions with participants.

\subsubsection*{Participants successfully used \name to accomplish writing tasks in ways that differ from traditional writing} This was highlighted by participants rating themselves 14\% more successful with \name. They also found \name more usable by rating it 21\% higher on the SUS. This trend was reflected in participants' comments. For example, P6 mentioned preferring \name because with the baseline, \textit{``it's what I'm using right now''} and \textit{``it's quite hard to adjust the prompts and I need [...] several rounds of prompting''}. P9 also mentioned that \name was \textit{``easier to use and more fun to navigate''} whereas with the baseline \textit{``you had to be very, very specific otherwise it would do something completely different''}. This confirms findings that people want more control when using LLMs~\cite{massonDirectGPTDirectManipulation2024, arnoldGenerativeModelsCan2021}.

\subsubsection*{\name was easier to use for many tasks, although some participants had different preferences} \label{sec:preferences}
Participants unanimously preferred \name overall and rated it as easier for most tasks. However, four participants expressed different preferences for some tasks, justifying their choice with personal preferences. For example, P3 preferred organizing text using the baseline because \textit{``putting [text] in the margin to me, felt like using a Figma board which I don't like''}. Similarly, P7 preferred exploring variations using the baseline because \textit{``I would see [the variations] in a separate interface instead of directly modifying my text''}. 
This points to the well-known design guideline of supporting ``many paths and many styles'', which strives to offer different ways of accomplishing tasks because people work in unpredictable ways~\cite{shneidermanCreativitySupportTools2006}.

\subsubsection*{Participants' knowledge of drawing software likely transferred to \name} The SUS question \textit{``'I would imagine that most people would learn to use this system very quickly.''} was the most highly rated. And a recurring comment from participants was how \name behaved like they would expect. P3 mentioned \textit{``the outcome was kind of what I expected would happen''}, P2 explained \textit{``some concepts are very [..] close to drawing software [...] if you've used it, you know how to use it [in Textoshop]''}, P9 also mentioned \textit{``some features the system has and the way it's organized, it's very similar to drawing software.. so... it's easier for me''}. In fact, P1 even commented on how they could guess how a tool would work before using it: \textit{``I didn’t really use the eyedropper, but I saw it and I could figure out what it’d do: pick up the tone of whatever sentence I placed it on''}.

\subsection{Observations and Participants' Comments}\label{sec:observations}

\subsubsection{The drawing analogy was mostly successful} Good interface analogies ``help establish user expectations and encourage predictions''~\cite{nealeChapter20Role1997}. We observed such expectations and predictions from our participants, as they mentioned how \name allowed a trial-and-error exploration~\cite{massonSuperchargingTrialErrorLearning2022} similar to the exploration they were used to in drawing software. P6 mentioned \textit{``when trying to pick colours for my own drawing, I usually randomly just adjust the point on the palette and just kind of see what I can get. So I adopted the same strategy when I was trying tones''} suggesting part of the success of the colour picker analogy is that it matched how one would use it in drawing software. 
However, one danger with interface analogies is the ``overly literal translation of existing bad designs''~\cite{preeceInteractionDesignHumanComputer2002}. We observed one instance of such phenomenon regarding the boolean operations, P1 explained \textit{``Even now with photo and image editing [software], ``unite'' is pretty straightforward, but all the other ones […] I often can’t remember which one means what''}. 
Finally, many participants described the system as ``fun'' and P3 commented \textit{``I thought it was fun just to play around with it [...] it kind of sparked that same joy that I get when I mess around with image editing software''}. This relates to the observation from \citet{carrollMetaphorCognitiveRepresentation1982} that an interface analogy that ``is conducive to the desired emotional attitude of the user'' and ``provide the user with exciting metaphors for routine work'' may enhance the intrinsic motivation of users in accomplishing their tasks.

\subsubsection{Participants quickly learned how to use \name, but it is unclear if they know when and how to integrate it into their workflow.} The user study tested the first part of the learning curve of \name, which corresponds to the time it takes for a user to reach a reasonable level~\cite{nielsenUsabilityEngineering1994}. Our results and observations suggest participants started high on the learning curve, possibly thanks to transferring their knowledge from drawing software~\cite{polsonTestCommonElements1986, nealeChapter20Role1997}. However, we only have anecdotal evidence about the shape of the second part of the learning curve, which corresponds to the transition from novice to expert. 
Specifically, we observed participants had no issues trying features and workflows and undoing them if necessary, which might help discover new features and expert workflows~\cite{massonSuperchargingTrialErrorLearning2022}. However, it remains unclear if people would actually integrate features of \name into their workflow. For example, it is known that, despite knowing more efficient ways of accomplishing tasks, such as keyboard shortcuts, users might still prefer their inferior approach~\cite{cockburnSupportingNoviceExpert2015, banovicTriggeringTriggersBurying2012}. Similarly, users and \name might co-adapt such that features of \name are appropriated and used in unexpected ways~\cite{mackayRespondingCognitiveOverload2000}.

\subsubsection{Identifying changes was sometimes challenging with \name} Some participants noted how a main difference with drawing software was that it is not as immediate to perceive textual changes. P8 commented \textit{``in like Figma [...] it's immediately, visually obvious what changed. [...] It didn't feel like I had the same mobility here where I could like immediately understand what the changes and the consequences were''}. In fact, some participants mentioned how they would prefer to have some time to review the changes before integrating them. P8 added \textit{``I really want to see things side-by-side''}. P6 explained that they would like a system to track the changes because the current animation of changes \textit{``disappeared so quickly before I can notice [the changes]''}. In \name, changes can be tracked either by working on copies or using layers, but these comments seem to indicate that, unlike drawing software, this is not enough to help users identify textual changes.

\subsubsection{\name forced pointing interactions} Unlike traditional text editors that can be used using only the keyboard, \name requires some pointing interaction to arrange text fragments, resize, rotate and manipulate tools. This was noted by both P7 and P8. P7 said \textit{``one difficulty is that [drawing software] is more heavily dependent on mouse or pen, so it's not about using the keyboard [..] one issue is switching between the two''}. P8 explained \textit{``I feel like text editing is inherently like a keyboard-driven task, and I felt like it was difficult to do that with [Textoshop]. [...] I think I just really know how to use the keyboard really quickly and fast''}.

\subsubsection{\name gave participants more control} Even though both \name and the baseline used the same underlying model, participants frequently mentioned control as an advantage of \name. For example, when referring to the tone picker, P4 explained \textit{``when I'm writing and I want a certain tone, it's nice to have a slider for that, because a lot of the times, text that ChatGPT generates is like weirdly wordy and formal. But then, when I ask make it casual, it makes it way too casual''}. When referring to the resizing feature, P11 explained \textit{``I don't want too much. The fact that I can just control like how much I want to elaborate on it, I think that was really cool and like, really useful''}.

\subsection{Limitations}
\subsubsection{\name might not be easy to learn for people with no experience with drawing software} While some participants reported not frequently using drawing software, all of them reported being familiar with drawing tools, colour pickers, layers, and boolean operations on shapes. We expect \name to lose its advantage in terms of learnability when used by people with little to no knowledge of drawing software.

\subsubsection{Due to the implementation of \name, it inherits all the issues of using LLMs to edit text} While the implementation of \name is not core to our contribution, it is important to acknowledge that \name currently heavily relies on OpenAI's GPT-4o. It is known that LLMs have issues in terms of biases, privacy, and plagiarism~\cite{weidingerEthicalSocialRisks2021, meyerChatGPTLargeLanguage2023}. We also found the model to perform poorly on tasks such as resizing to a specific number of words. In our implementation, we workaround these issues by regenerating multiple versions until one that matches the length is found. But it is often imperfect, which might frustrate users.
Finally, some text transformations might result in imperfect, incongruent, or hallucinated text. These issues might be mitigated by using fine-tuned LLMs that take greater care in preserving semantic meaning and matching the style of the original text.

\subsubsection{The drawing software analogy might confuse users} Using interface analogies might lead to mismatches between the source and the target domain. While some mismatches might be beneficial and lead to curiosity, others might cause problems and create invalid assumptions~\cite{nealeChapter20Role1997}. With \name, this difficulty is coupled with the use of an LLM that might produce unexpected text, making the system difficult to model for the user~\cite{chungIntersectionUsersRoles2021}. While this did not seem to be an issue during the study, some participants did mention a mismatch between their mental model and the behaviour of the system regarding the layer system. For example, P3 mentioned not seeing text layers like stacked sheets but more like \textit{``variables or conditions that you're kind of swapping in and out''}. One possible explanation is given by \citet{nealeChapter20Role1997} in that ``mismatches occur with aspects of computer functionality when the target domain is unable to provide a direct real-world correlate''. Text layers have no real-world equivalent. Unlike physical transparent sheets, text layers essentially replace text and are positioned dynamically based on their anchor points in the text. Finally, text layers also led users to make invalid assumptions. For example, P8 expected features and were surprised not to find them \textit{``In Figma [...] when I click again it goes down a layer rather than staying at the same level''}

\subsection{Future Work}

\subsubsection{Repurposing other digital drawing features} We only scratched the surface of drawing features that could be repurposed for text. For example, layers could support different blending modes to either overwrite or track the changes and changing the opacity of the layer could control how the text is integrated (e.g., 100\% overwrites, while 50\% generates text with a meaning midway between the layer and what it overlaps). Other layer types could include ``Adjustment layers''~\cite{andrewsAdobePhotoshopElements2006} that apply a transformation to the text (e.g., making the text more positive) while retaining the integrity of the text underneath. The rotation feature could allow specifying the centre of rotation (the word that should remain at the same location). Finally, text could be transformed via histograms showing text properties such as the number of adverbs, the sentiment of the text, or the length of the sentences. Then, writers could manipulate the histogram to transform the text.

\subsubsection{Leveraging recent research in graphical editing to also benefit \name} The design of drawing software is a well-researched topic. Much of this research could be adapted and implemented in \name. For example, work on reifying~\cite{beaudouin-lafonReificationPolymorphismReuse2000} graphical attributes into manipulable objects~\cite{xiaObjectOrientedDrawing2016} could be adapted so that properties such as tone, sentence length, and sentence rotation become first-class objects that can be easily manipulated, copied, reversed, and removed. Work on better colour pickers and colour theme authoring could similarly power better tone pickers where tones become objects that can be rearranged and interpolated~\cite{shugrinaColorBuilderDirect2019}. Similarly, work on using histograms as another editing space could inform the design of text histograms that can query and edit the text~\cite{chevalierHistomagesFullySynchronized2012}.

\subsubsection{Testing the impact of \name over a longer period of time} 
While challenging due to the nondeterministic nature of LLMs that might confound the results, a longitudinal or diary study of writers using \name over several months would tell us how people integrate the features in their workflows, how they co-adapt with the system~\cite{mackayRespondingCognitiveOverload2000}, and how it impacts the ideation, the quality of the result, and the productivity~\cite{chungIntersectionUsersRoles2021}.
For example, the consequences of constantly switching between keyboard and pointing devices are unclear. Moving from the keyboard to the mouse takes about 0.4s~\cite{cardEvaluationMouseRateControlled1978} and might have a cognitive cost which might add up in the long run. 
However, we note that these switches may not be so frequent considering writing is not only about typing characters on a keyboard but also involves planning and reviewing, which do not leverage the keyboard~\cite{flowerCognitiveProcessTheory1981}. Additionally, \citet{buschekCollageNewWriting2024} observed a shift in how writing is done to look more like an assemblage of text snippets rather than manual writing. Such workflow is less reliant on the keyboard and might be better supported by \name.
Beyond productivity, a longitudinal study would also help study how \name changes the writing workflow and the text that is produced. For example, the diverse ways participants used layers to structure their text is reminiscent of fluid documents~\cite{changFluidlyRevealingInformation2000} where the added structure and ways to reveal in-context information changed people's reading behaviours~\cite{zellwegerImpactFluidDocuments2000}. While some features of \name may not make writing easier nor faster, they might contribute to other aspects such as idea generation or quality of the craft~\cite{chungIntersectionUsersRoles2021}.

\subsubsection{Applying the drawing software analogy to other domains} Analogies used to design software often stay within the same domain: the digital canvas for drawing software and the blank sheet for text editors. Through this work, we show the potential of using analogies from different domains. We believe this strategy could inform the design of software from all domains. For example, code editors could also benefit from layers to structure the code. This would allow an organization with a finer granularity than files, and layers could be hidden to remove the code responsible for a specific feature. Algorithms could be merged, intersected or excluded to change their behaviours. And tools such as brushes could perform smart editing of code to refactor, add comments, or change the coding style. In fact, this idea was recently experimented with ``Code Brushes'' that can enhance the readability or fix bugs in code~\cite{wattenbergerGitHubNextCode2024}.

\section{Conclusion}
We presented \name, a text editor that repurposes features from drawing software to offer new workflows and features that facilitate text editing. Results of a user study showed that participants were more successful and preferred \name to accomplish text editing tasks. The study also highlighted how participants transferred their knowledge of drawing software to learn how to use \name. Broadly, our work highlights the potential of software analogies to inform the design of software from any domain.

\begin{acks}
This work was supported by NAVER corporation as a part of NAVER-Wattpad-University of Toronto research center and NSERC Discovery Grant RGPIN-2018-05072.
\end{acks}




\bibliographystyle{ACM-Reference-Format}
\bibliography{_references.bib, zotero_do_not_modify}




\end{document}